\pdfoutput=1

\documentclass[11pt]{article}

\usepackage[preprint]{acl}

\usepackage{times}
\usepackage{latexsym}
\usepackage[T1]{fontenc}
\usepackage[utf8]{inputenc}
\usepackage{microtype}
\usepackage{inconsolata}
\usepackage{graphicx}
\usepackage{amsmath} 
\usepackage{algorithm}
\usepackage{algorithmic}
\usepackage{float}
\usepackage{booktabs}    
\usepackage{multirow}    
\usepackage{array}       
\usepackage{amsfonts}  
\usepackage{amssymb}   
\usepackage{tabularx}   
\usepackage{makecell}   
\usepackage{tcolorbox}

\newcommand{\fitwide}[1]{%
  \begingroup
  \setlength{\tabcolsep}{3pt}
  \resizebox{\textwidth}{!}{#1}
  \endgroup
}

\newcommand{\Autoref}[1]{%
  \begingroup%
  \def\chapterautorefname{Chapter}%
  \def\sectionautorefname{Section}%
  \def\subsectionautorefname{Subsection}%
  \def\subsubsectionautorefname{Subsubsection}%
  \def\paragraphautorefname{Paragraph}%
  \def\tableautorefname{Table}%
  \def\equationautorefname{Equation}%
  \def\algorithmautorefname{Algorithm}%
  \autoref{#1}%
  \endgroup%
}

\title{Multimodal Item Scoring for Natural Language Recommendation via Gaussian Process Regression with LLM Relevance Judgments}
\author{Yifan Liu\thanks{Equal contribution}\textsuperscript{\textnormal{1}},
  Qianfeng Wen\footnotemark[1]\textsuperscript{\textnormal{1}}, Jiazhou Liang\footnotemark[1]\textsuperscript{\textnormal{1}}, Mark Zhao\footnotemark[1]\textsuperscript{\textnormal{1}},\\ \textbf{Justin Cui\textsuperscript{\textnormal{1}}, Anton Korikov\textsuperscript{\textnormal{1}}, Armin Toroghi\textsuperscript{\textnormal{1}}, Junyoung Kim\textsuperscript{\textnormal{2}}
  and Scott Sanner\textsuperscript{\textnormal{1}}}\\
  \textsuperscript{1}University of Toronto, Canada\\
  \textsuperscript{2}Sungkyunkwan University, South Korea \\
  \texttt{yifanliu.liu@mail.utoronto.ca}\\
}

\begin{document}
\maketitle

\begin{abstract}
  Natural Language Recommendation (NLRec) generates item suggestions based on the relevance between user-issued NL requests and NL item description passages. Existing NLRec approaches often use Dense Retrieval (DR) to compute item relevance scores from aggregation of inner products between user request embeddings and relevant passage embeddings. However, DR views the request as the sole relevance label, thus leading to a unimodal scoring function centered on the query embedding that is often a weak proxy for query relevance.  To better capture the potential multimodal distribution of the relevance scoring function that may arise from complex NLRec data, we propose \textbf{GPR-LLM} that uses Gaussian Process Regression (GPR) with LLM relevance judgments for a subset of candidate passages. Experiments on four NLRec datasets and two LLM backbones demonstrate that GPR-LLM with an RBF kernel, capable of modeling multimodal relevance scoring functions, consistently outperforms simpler unimodal kernels (dot product, cosine similarity), as well as baseline methods including DR, cross-encoder, and pointwise LLM-based relevance scoring by up to 65\%. Overall, GPR-LLM provides an efficient and effective approach to NLRec within a minimal LLM labeling budget.
\end{abstract}

\section{Introduction}
\label{sec:intro}

\begin{figure}
    \centering
    \includegraphics[width=0.91\linewidth]{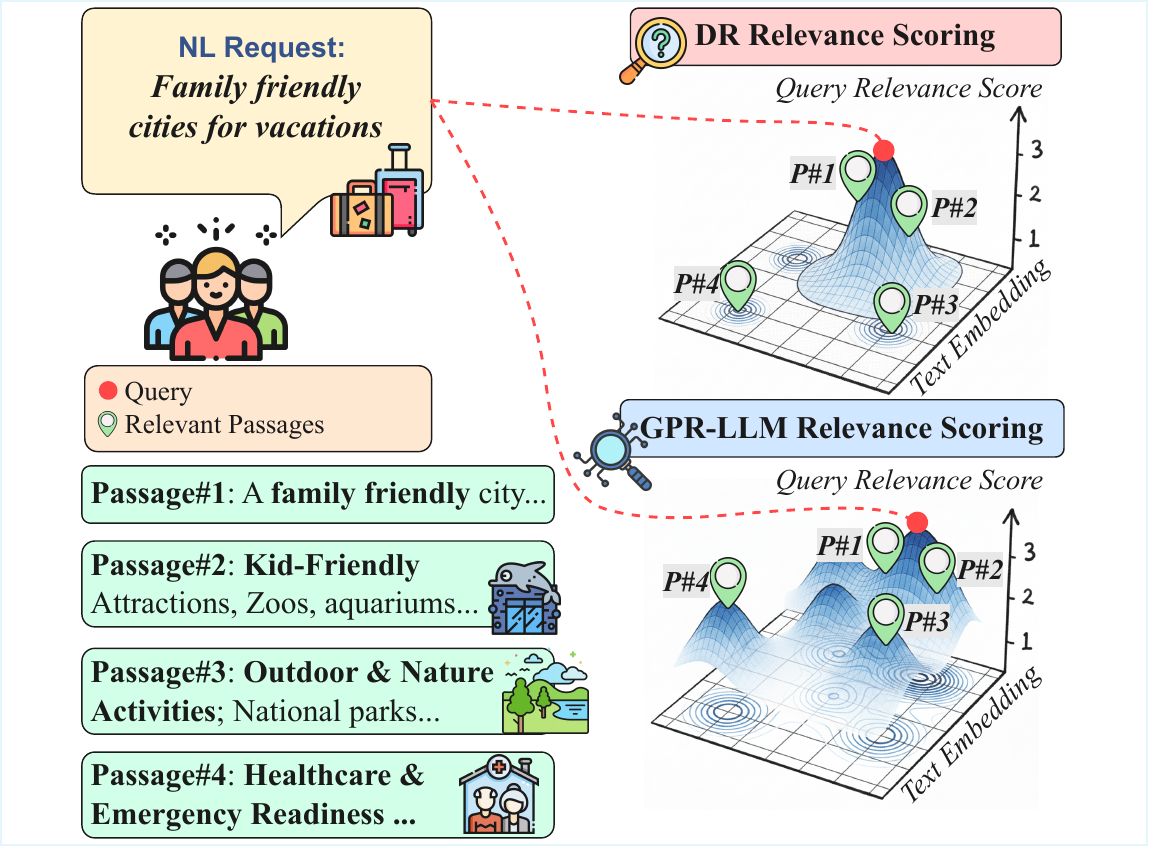}
    \caption{
    \textbf{(Top)} Dense retrieval (DR) methods assume a unimodal relevance scoring function concentrated near the user's NL request within the embedding space. \textbf{(Bottom)} In practice, the true relevance function is often multimodal, with relevant passages dispersed across distinct regions. The relevance scoring function for NL request "\textit{Family friendly cities for vacations}" is multimodal as it covers multiple theme passages, including "\textit{Kid-friendly Attractions}," "\textit{Outdoor and Nature Activities}," and "\textit{Healthcare and Emergency Readiness}." 
    }
    \label{fig:multimodal-dense}
\end{figure}

Natural Language Recommendation (NLRec) \cite{Kang2017} aimsGuassianto generate item suggestions based on user-issued free-form textual NL requests. Unlike traditional recommender systems that rely on historical interaction data, NLRec assumes the request itself encodes the user’s preferences and intent \cite{Kang2017, ndr2017}. Each item is typically associated with multiple descriptive passages such as summaries, reviews, or menus. To score item relevance, existing approaches commonly use Dense Retrieval (DR; \citealp{karpukhin2020dense}), which first estimates passage relevance scores based on cosine similarity or inner product between NL request and passage embeddings in a dense embedding space, and then aggregates these passage-level scores into item-level scores. The standard DR approach implicitly assumes that the relevance scoring function is unimodal with a peak centered around the NL request (\autoref{fig:multimodal-dense}, top). However, in practice, relevance is often complex and multimodal: the NL request alone is insufficient as the sole positive relevance label. Passages may require complex reasoning to determine relevance and may thus be distributed across multiple distant regions (modes) within the embedding space (\Autoref{fig:multimodal-dense}, bottom). As a result, the standard DR relevance scoring function is a weak proxy for passage-level relevance, and hence aggregate item relevance.


\begin{figure*}
    \centering
    \includegraphics[width=1.0\linewidth]{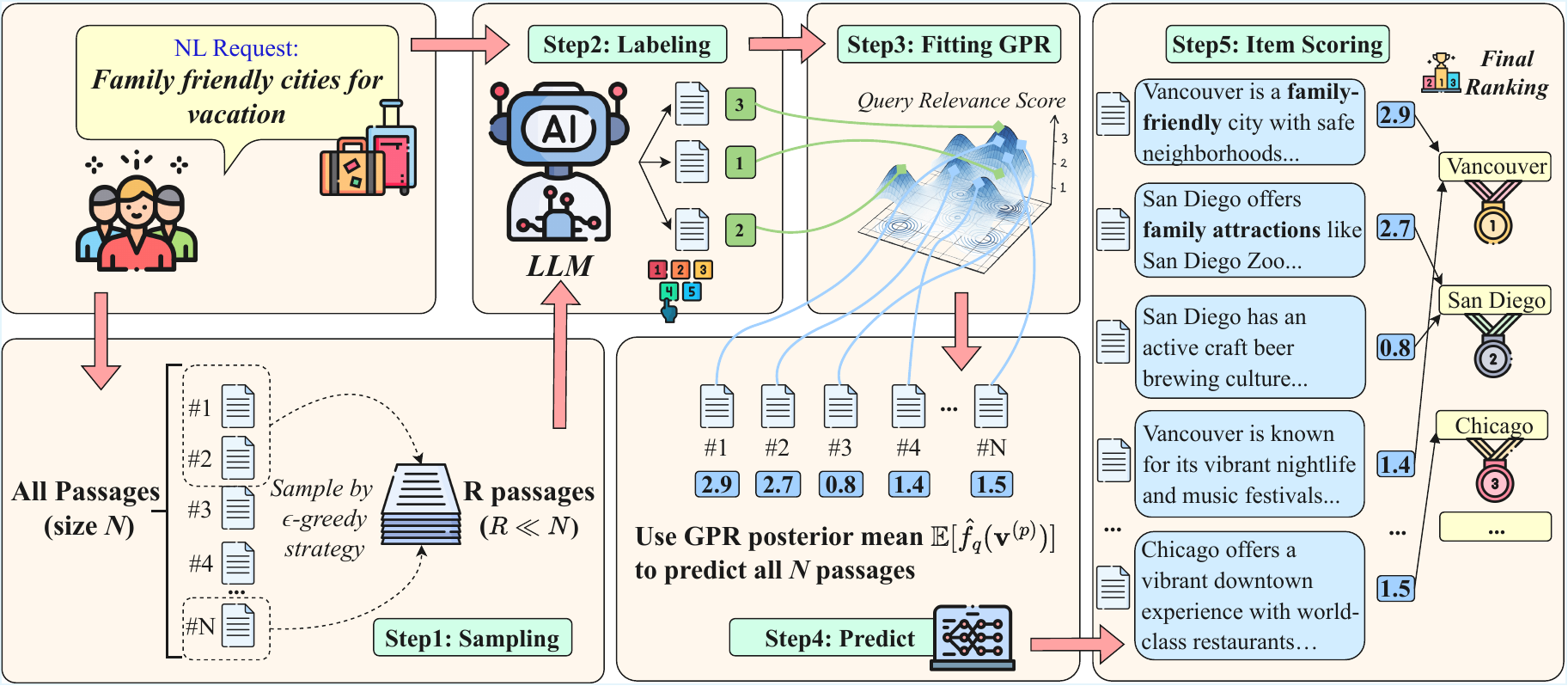}
    \caption{Overview of GPR-LLM. A small subset of $R \ll N$ passages is sampled from the full set of $N$ passages using an $\epsilon$-greedy sampling strategy and labeled using an LLM. A query-specific GPR model is then trained on this labeled subset to estimate relevance scores for \emph{all} passages. Finally, item-level relevance scores are aggregated from the passage-level scores using \autoref{eq:item-level}, and items are ranked accordingly to generate the recommendation list.}
    \label{fig:Pipline}
\end{figure*}

Large Language Models (LLMs) have recently emerged as a promising resource for reasoning about relevance between items or passages and NL queries, which offer more reliable relevance scoring than can be achieved with DR \cite{LLM:qg, LLM:prp, LLM:rankgpt, LLM:zslistwise, LLM:yesno, LLM:zeroretriever}. However, exhaustively prompting LLMs for relevance judgments of all passages associated with an item is expensive, especially with large item collections and numerous NL description passages.

In order to improve the reliability of the dense scoring function for NLRec while working within a minimal budget of LLM labeling calls, we propose \textbf{GPR-LLM} that uses Gaussian Process Regression (GPR) \cite{rasmussen2006gp, gramacy2020surrogates} with LLM relevance judgements for a sampled subset of candidate passages. 

A key insight of GPR-LLM is that the similarity or distance functions from DR can serve directly as \textit{kernel functions} with GPR. In the special case where the kernel function is identical to the similarity function used in DR, and assuming that the NL request itself is the only relevant label, GPR exactly reduces to standard DR. However, GPR can also critically leverage \emph{additional relevance labels} obtained from LLM passage relevance judgments to estimate the multimodal relevance scoring function (cf. \autoref{fig:multimodal-dense}, bottom) when used with an RBF kernel over dense embeddings. As we show empirically, GPR-LLM is an efficient and effective relevance scoring approach for NLRec that outperforms state-of-the-art methods under a minimal LLM labeling budget.
In summary, we make the following key contributions: \begin{itemize} 

\item We propose GPR-LLM (cf. \Autoref{fig:Pipline}) to estimate the multimodal relevance scoring function in NLRec using GPR with LLM relevance judgements to address the limitations of the unimodal relevance scoring function in DR. 

\item We empirically show that the RBF kernel, which is designed to capture the multimodal relevance scoring function, consistently outperforms unimodal dot product and cosine similarity kernels across four benchmarks and two LLM backbones.

\item 
GPR-LLM consistently outperforms all baselines, including BM25, DR, Cross-encoder, and LLM-based Relevance Scoring under the same LLM labeling budget, and achieves comparable performance to the best baselines even with a significantly smaller budget.
\end{itemize}

\section{Problem Definition}
We consider the problem of ranking a set of items in response to a user-issued NL request, which we simply refer to as a query \( q \in \mathcal{Q} \). Let \( \mathcal{I} \) represent a collection of total \( M \) items. Each item \( i \in \mathcal{I} \) is associated with a set of textual passages. We denote the complete collection of \( N \) passages as \( \mathcal{P} \), and their corresponding text embeddings as \( \mathcal{V} \), where \( \mathbf{v}^{(p_j)} \in \mathbb{R}^D \) represents the embedding of passage \( p_j \in \mathcal{P} \).
Each passage \( p_j \in \mathcal{P} \) is uniquely linked to item \( i \).  We define the set of item \( i \)'s passages as
\begin{equation}
    \mathcal{P}^{(i)} = \{ p_j \in \mathcal{P} \mid p_j \text{ is associated with item } i \}.
\end{equation}



The objective is to recommend the item \( i \in \mathcal{I} \) that is most relevant to \( q \) based on its associated passages \( \mathcal{P}^{(i)} \). Thus, we decompose the task into two subproblems: passage-level relevance estimation and item-level relevance aggregation.

\paragraph{Passage-Level Relevance}
Let 
$
f^*_q : \mathcal{P} \to \mathbb{R}
$
be a query-specific passage relevance scoring function
that assigns a ground truth real-valued relevance score to passage \( p_j \in \mathcal{P} \) with respect to query \( q \). 
let \( \mathcal{S}^* \in \mathbb{R}^{M \times N}\) be the ground truth relevance score of the $j$-th passage of item \( i \)as:
\begin{equation}
\mathcal{S}^*_{i,j} = f^*_q(p_j) \quad \forall p_j \in \mathcal{P}^{(i)} , i \in \mathcal{I}
\end{equation}
However, in real-world applications, \( f^*_q(p_j) \) is often unknown; thus, we employ $\hat{f}_q : \mathcal{P} \to \mathbb{R}$ to estimate $\mathcal{S}^*$ as $\mathcal{S}$ where $\mathcal{S}_{i,j} = \hat{f}_q(p_j) \forall p_j \in \mathcal{P}^{(i)}$.

\paragraph{Item-Level Relevance}
Given the estimated passage-level relevance score $\mathcal{S}_{:,:}$, item-level scores can be derived by aggregating  $\mathcal{S}_{:,:}$ by item. Specifically, for each item \( i \), we first select the top-\( T \) passages from \( \mathcal{P}^{(i)} \) with highest \( \mathcal{S}_{i,:} \). We denote this list as \( \mathcal{P}^{(i)}_{\text{top-T}} \).
The item-level relevance score \( \mathcal{S}_i \) for an item \( i \) is computed as follows:
\begin{equation}\label{eq:item-level}
    \mathcal{S}_i = \phi\left(\left[\, \mathcal{S}_{i,j} \mid p_j \in \mathcal{P}^{(i)}_{\text{top-T}} \,\right]\right) \quad\forall  i \in \mathcal{I},
\end{equation}
where \( \phi: \mathbb{R}^T \rightarrow \mathbb{R} \) is an aggregation function (e.g., mean or max).
The top-\( K \) NLRec items are ranked by $\mathcal{S}_i$ in descending order.

The effectiveness of NLRec relies heavily on the quality of estimated passage-level scores. Thus, we focus on the formulation of \( \hat{f}_q \), aiming to closely approximate the true relevance function \( f^*_q \) by effectively capturing the multimodal relevance scoring function, thereby enabling better item-level relevance scoring estimation for NLRec.

\section{GPR-LLM: Gaussian Process Regression with LLM Relevance Judgments}

To capture the multimodal structure of \( \mathcal{S}^* \) (cf. \Autoref{fig:multimodal-dense}, bottom) and ensure generalizability across different applications, we aim to approximate the true (possibly multimodal) relevance function \(f_q^*\) with a query-specific GPR scorer \(\hat{f}_q:\mathcal{V}\to\mathbb{R}\) that leverages both (i) LLM relevance judgments on a small, carefully sampled subset of passages and (ii) the geometry of NL embedding spaces (as in DR). At a high level:
\begin{enumerate}
    \item \textit{Candidate Sampling:} from all \(N\) passages, construct a small labeled set \(\mathcal{P}^{\mathrm{GP}}\) of size \(R\ll N\) via an \(\epsilon\)-greedy strategy that mixes top DR-ranked items and uniform random exploration (cf. \Autoref{sec:sample}).
    
    \item \textit{LLM Relevance Judgments:} Obtain \(\mathcal{S}_{\cdot,j}^{\mathrm{LLM}}\) from an LLM backbone for each sampled passage \(p_j \in \mathcal{P}^{\mathrm{GP}}\) (cf.~\Autoref{sec:llm_relevance}).

    \item \textit{Fitting GPR:} fit a query-specific GP prior/posterior over \(\hat{f}_q\) with a kernel defined in the embedding space (cf.~\Autoref{sec:gpr_spec} and~\Autoref{sec:kernel}).
    
    \item \textit{Scoring all passages:} use the GPR posterior mean \(\mathbb{E}[\hat{f}_q(\mathbf{v}^{(p)})]\) to produce passage-level scores for all \(p\in\mathcal{P}\), followed by standard item-level aggregation.
\end{enumerate}

\subsection{Query-specific GPR.}
\label{sec:gpr_spec}
Gaussian Process Regression (GPR) \cite{Williams1996} is a non-parametric, kernel-based regression model that has been widely adopted for estimating multimodal functions \cite{Lawrence2009}.

GPR places a GP prior over \(\hat{f}_q\):
\begin{equation}  \label{eq:gp_pior}
    \hat{f}_q(\mathcal{V}) \sim \mathcal{GP}(0, k(\mathcal{V}, \mathcal{V}^\prime)),
\end{equation}
where \( k(x, x') \) is a covariance function (kernel). 

Given \(R\) passages with observed scores \(\mathcal{S}_{\cdot,j}\) and the query with max possible score, we collect scores and embeddings as
\begin{equation}\label{eq:D}
\mathcal{D} = \{(\mathbf{v}^{(q)}, \mathcal{S}_{\max})\}\cup \{(\mathbf{v}^{(p_j)}, \mathcal{S}_{\cdot,j})\}_{j=1}^R.
\end{equation}
For a passage \( p_* \) with embedding \( \mathbf{v}^{(p_*)}\), the noisy-observation posterior is
\(p(\hat{f}_q(\mathbf{v}^{(p_*)})|\mathcal{D}) \sim \mathcal{N}(\boldsymbol{\mu}_*, \boldsymbol{\Sigma}_*)\) with
\begin{equation}\label{eq:e}
    \boldsymbol{\mu}_* = \mathbf{k}_*^{\top} (\mathbf{K} + \alpha \mathbf{I})^{-1} \{\mathcal{S}_{\cdot,j}\}_{j=1}^R,
\end{equation}
\begin{equation}\label{eq:var}
    \boldsymbol{\Sigma}_*= k(\mathbf{v}^{(p_*)}, \mathbf{v}^{(p_*)}) - \mathbf{k}_*^{\top} (\mathbf{K} + \alpha \mathbf{I})^{-1} \mathbf{k}_*,
\end{equation}
where
\begin{itemize}
  \item \( \mathbf{K} \in \mathbb{R}^{R \times R} = k(\{\mathbf{v}^{(p_j)}\}_{j=1}^R, \{\mathbf{v}^{(p_j)}\}_{j=1}^R) \) is the kernel matrix of the labeled passages,
  \item  \( \mathbf{k}_* \in \mathbb{R}^R = k(\{\mathbf{v}^{(p_*)}, \{\mathbf{v}^{(p_j)}\}_{j=1}^R)\) is the cross-kernel vector, and
  \item \( \alpha \) is the variance of the Gaussian noise (label noise) of observed $\{\mathcal{S}_{\cdot,j}\}_{j=1}^R$ since the true relevance is usually unknown. 
\end{itemize}
The predictive mean \( \mathbb{E}[\hat{f}_q(\mathbf{v}^{(p_*)})] \) is used as the estimated passage relevance, which we later aggregate to the item level.

\subsection{Kernels for GPR} \label{sec:kernel} 
 We can conveniently repurpose the distance or similarity function used in DR as a kernel function for GPR to leverage the intuition of DR that semantically similar NL texts are likely to be located close together in the embedding space. We propose the following dense kernels:

\begin{itemize}
    \item \textbf{Dot Product (Linear)}: $k(\mathcal{V}, \mathcal{V}') = \mathcal{V}^\top \mathcal{V}'$
    \item \textbf{Cosine Similarity}:
    $k(\mathcal{V}, \mathcal{V}') = \frac{\mathcal{V}^\top \mathcal{V}'}{\|\mathcal{V}\| \cdot \|\mathcal{V}'\|}$
    \item \textbf{Radial Basis Function (RBF)}: \newline
            $k(\mathcal{V}, \mathcal{V}') = \exp\!\big(-\frac{\|\mathcal{V} - \mathcal{V}'\|^2}{2\ell^2}\big)$
\end{itemize}
Both the Dot Product and Cosine Similarity kernels are consistent with the relevance measures used in DR. They are \textit{non-stationary} kernels capable of modeling only unimodal distributions~\cite{williams2006gaussian}. In contrast, RBF is a \textit{stationary} kernel that measures similarity by proximity in the embedding space and capture local variations across regions. Due to this locality~\cite{steinwart2001influence, williams2006gaussian}, it is effective for capturing the potentially multimodal relevance function (cf.~\Autoref{fig:multimodal-dense}, bottom). 

\subsection{LLM Relevance Judgments for GPR}
\label{sec:llm_relevance}

Another crucial consideration for GPR-LLM is the source of relevance judgements $\mathcal{S}$ in $\mathcal{D}$. LLMs are capable of modeling complex relationships between queries and passages that require contextual reasoning, inference, or multi-step understanding \cite{LLM:rankgpt,LLM:prp}. This makes LLM relevance judgments a natural choice to estimate $S^*$.

We use the commonly adopted UMBRELA \texttt{prompt}~\cite{Upadhyay2024umb} and follow the procedure of \citet{LLM:yesno} to obtain the LLM relevance judgment between a query $q$ and a passage $p_j$ as follows:
\begin{equation}
    \label{eq:llm_prompt}
    \mathbf{z} = \mathrm{LLM}(q, p_j, \texttt{prompt})
\end{equation}
where $\mathbf{z} = [z_0, z_1, \dots, z_{K-1}]$ corresponds to the logit of a predefined discrete relevance label $r_k \in \{0, 1, \dots, K-1\}$.  
We then compute the LLM-based relevance score using the \textit{expected relevance}:
\begin{equation}
    \label{eq:llm_er}
    \mathcal{S}_{i,j}^{\mathrm{LLM}} = \sum_{k=0}^{K-1} \left( \frac{e^{z_k}}{\sum_{j=0}^{K-1} e^{z_j}} \right) \cdot r_k.
\end{equation}

\subsection{Sampling for GPR}
\label{sec:sample}

Since obtaining LLM relevance judgments of the entire $N$ passages is expensive, we instead sample a subset of passages (i.e., \( R \ll N\)), denoted as 
\(
    \mathcal{P}^{\mathrm{GP}}
\).
In an ideal situation, $\mathcal{P}^{\mathrm{GP}}$ should include all relevant passages. DR offers an efficient mechanism for identifying potentially relevant passages. However, its underlying unimodal assumption often causes relevant passages far from the query in the embedding space to be overlooked.

To mitigate this limitation, we propose the \textit{\(\epsilon\)-Greedy Sampling} method, which leverages DR efficiency while reducing its inherent bias. Specifically, we define the sampling set as:
\[
\mathcal{P}^{\mathrm{GP}} = \mathcal{P}^{\mathrm{DenseRanking}} \;\cup\; \mathcal{P}^{\mathrm{ExploratorySampling}},
\]
where \(\mathcal{P}^{\mathrm{DenseRanking}}\) consists of the top-\(\lfloor(1-\epsilon)R\rfloor\) passages ranked by DR. The remaining \(\lceil\epsilon R\rceil\) passages are sampled uniformly at random from the top-\(\eta\) DR-ranked passages not already selected, where \(\eta \gg \lfloor(1-\epsilon)R\rfloor\). Formally, 
\[
\mathcal{P}^{\mathrm{ExploratorySampling}} \subseteq \mathcal{P}^{\eta}\setminus\mathcal{P}^{\mathrm{DenseRanking}},
\]
with \(\mathcal{P}^{\eta}\) denoting a set of top-\(\eta\) DR-ranked passages. The method thus involves two parameters:
\begin{description}
    \item[]$\epsilon \in [0,1]$ controls the exploration–exploitation trade-off. Setting \( \epsilon = 0 \) corresponds to \textit{greedy sampling}, relying on DR rankings, whereas \( \epsilon = 1 \) selects passages uniformly at random.
    \item[]\( \eta \) constraints exploratory sampling to the top-\(\eta\) ranked passages, preventing the inclusion of irrelevant passages ranked too low by DR.
\end{description}

\( \mathcal{D} \) combines the LLM relevance judgments \( \mathcal{S}_{\cdot,j}^{\mathrm{LLM}} \) and the dense embedding \( \mathbf{v}^{(p_j)}\) of all sampled \( p_j \in \mathcal{P}^{\mathrm{GP}}  \) using \(\epsilon\)-Greedy Sampling. We then estimates the passage-level scores for all $\mathcal{P}$ given \( \mathcal{D} \) (cf. \Autoref{eq:e}) and aggregates for item recommendation (cf. \Autoref{eq:item-level}). \Autoref{fig:Pipline} illustrates the overall GPR-LLM pipeline.





\subsection{Complexity Analysis}

Per query, candidate sampling via $\epsilon$-greedy selection costs $\mathcal{O}(ND)$, corresponding to a dense-retrieval pass over $N$ passages with $D$-dimensional embeddings and negligible additional sampling overhead. LLM relevance judgments for $R$ passages incur a cost of $\mathcal{O}(R\,C_{\text{LLM}})$, where $C_{\text{LLM}}$ denotes the per-call inference cost. GPR fitting then includes kernel construction of $\mathcal{O}(R^2D)$ and an exact Cholesky decomposition of $\mathcal{O}(R^3)$. Finally, scoring all $N$ passages requires $\mathcal{O}(NRD)$ to compute the cross-kernel $K_{N\times R}$ and posterior mean. The total per-query runtime is therefore:
\[
\mathcal{O}\!\big(ND + R\,C_{\text{LLM}} + R^2D + R^3 + NRD\big).
\]
Since $R \!\ll\! N$, the dominant costs are typically the DR pass and GP inference over all passages, while GPR fitting remains inexpensive. In practice, the computational overhead of GPR-LLM relative to pointwise LLM-based relevance scoring with the same $R$ is minor (cf.~\autoref{tab:complexity_latency}).


\begin{table}[t]
\centering
\scriptsize
\begin{tabularx}{\columnwidth}{l X c}
\toprule
\textbf{Method} & \textbf{Per-query Complexity} & \textbf{Latency (sec)} \\
\midrule

\textbf{DR} & $\mathcal{O}(ND)$ & 0.165 (0.161, 0.168) \\
\cmidrule(lr){1-3}
\addlinespace[1pt]
\multicolumn{3}{c}{\textbf{LLM-based Scoring}}\\
\textbf{Point.} & $\mathcal{O}(ND + R\,C_{\text{LLM}})$ & 0.678 (0.671, 0.685) \\
\cmidrule(lr){1-3}

\addlinespace[1pt]
\multicolumn{3}{c}{\textbf{GPR\textendash LLM}}\\
\textbf{Dot} &
\multirow{3}{*}{\parbox[t]{0.74\linewidth}{%
$\mathcal{O}\big(ND + NRD\ + R\,C_{\text{LLM}} + R^2D + R^3\big)$}} 
& 0.782 (0.769, 0.795) \\
\textbf{Cosine} &  & 0.774 (0.762, 0.787) \\
\textbf{RBF}    &  & 0.754 (0.730, 0.780) \\
\bottomrule
\end{tabularx}
\caption{
Per-query time complexity and latency (seconds) under $R{=}50$, $N{=}100{,}000$, and $D{=}368$. Latencies show 95\% CIs in $[\cdot]$. All computations were performed on an NVIDIA GeForce RTX 4070 GPU.
}
\label{tab:complexity_latency}
\end{table}

\section{Experiments}

\subsection{Experimental Setup}

\begin{figure*}[htbp]
    \centering
    \includegraphics[width=\linewidth]{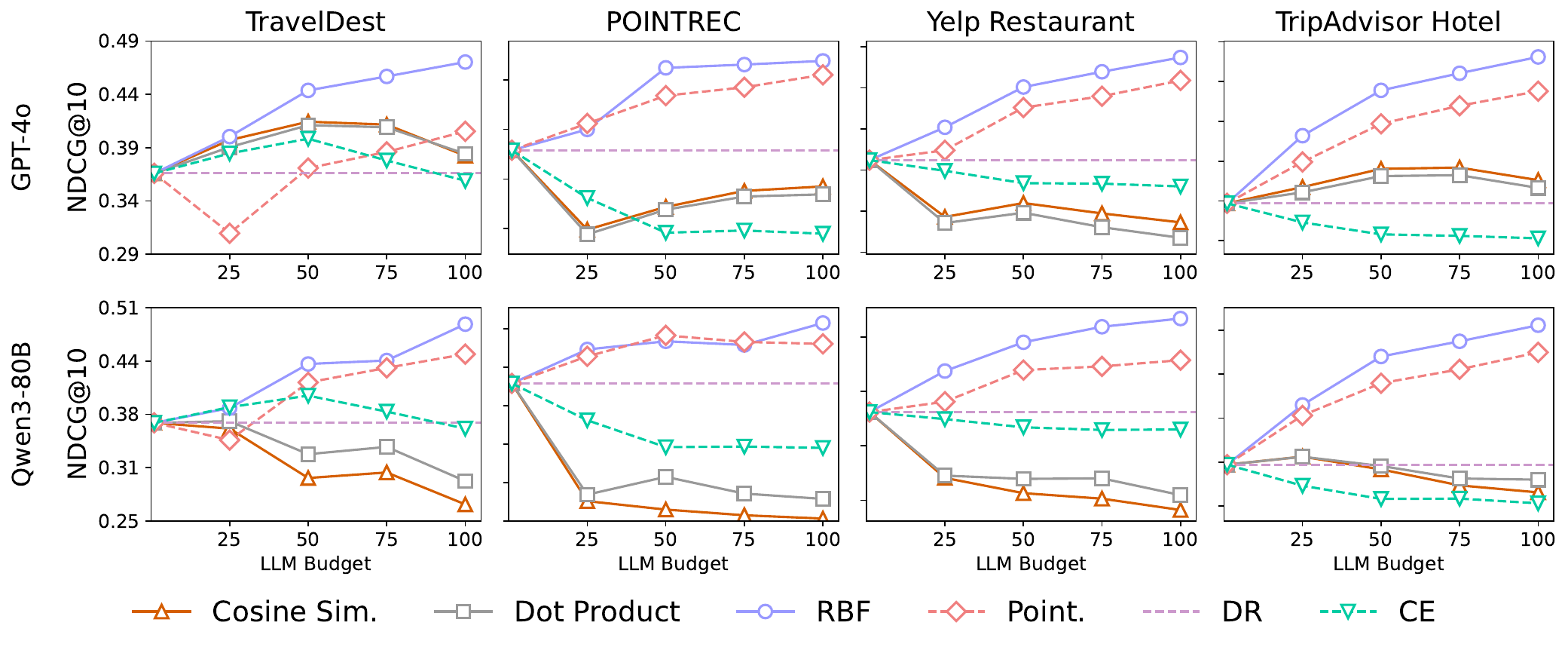}
    \caption{\textbf{RQ1:} Kernel choices comparison with Greedy Sampling ($\epsilon = 0$) under varying LLM labeling budget.
    }
    \label{fig:kernel-comparison-rq1}
\end{figure*}

We compare GPR-LLM against the following baseline relevance scoring methods\footnote{Code for reproducing our experiments is available at \url{https://anonymous.4open.science/r/Bandit_Retrieval-0352/README.md}.}\footnote{See \autoref{sec:impl_details} for detailed baseline implementations.}:

\begin{itemize}
    \item BM25~\cite{bm25}: Computes relevance scores using BM25.
    \item Dense Retrieval (DR)~\cite{karpukhin2020dense}: Computes relevance scores using inner products between query and passage embeddings.
    \item Cross-Encoder (CE)~\cite{nogueira2019multistage}: Computes relevance scores by jointly encoding query-passage pairs.
    \item Pointwise LLM-based Relevance Scoring (Point.)~\cite{LLM:yesno}: Computes relevance scores by prompting an LLM with query–passage pairs and computing the expected relevance score from the model’s predicted label distribution.

\end{itemize}

We conduct experiments on four publicly available benchmark datasets:
\begin{itemize}
    \item \texttt{POINTREC}~\cite{afzali2023pointrec}: Point-of-interest recommendation using reviews and structured descriptions.
    \item \texttt{TravelDest}~\cite{Wen2024}: Travel city recommendation using city-level descriptions.
    \item \texttt{TripAdvisor Hotel}~\cite{wen2025}: Hotel recommendation using user reviews.
    \item \texttt{Yelp Restaurant}~\cite{wen2025}: Restaurant recommendation using user reviews.
\end{itemize}

These datasets allow us to assess the generalizability of our approach across diverse NLRec scenarios. See \Autoref{tab:datasets} for dataset statistics.

\begin{table}[t]
    \centering
    \scriptsize
    \begin{tabularx}{\columnwidth}{lcccc}
        \toprule
        \textbf{Dataset} & \textbf{\# Queries} & \textbf{\# Items} & \textbf{\# Passages} & \textbf{\# Qrel} \\
        \midrule
        TravelDest & 50 & 775 & 126,400 & 3,721\\
        POINTREC & 28 & 59,553 & 592,818 & 498\\
        TripAdvisor Hotel & 100 & 589 & 133,759 & 4,887\\
        Yelp Restaurant & 100 & 1,152 & 283,658 & 11,726\\
        \bottomrule
    \end{tabularx}%
    \caption{Statistics of our benchmark datasets. Qrels indicates the total number of relevant items summed over all queries.}

    \label{tab:datasets}
\end{table}

We experiment with two LLM backbone models:
\begin{itemize}
    \item GPT-4o \cite{openai2023gpt4}
    \item Qwen3-Next-80B-A3B as an open-source model \cite{imple:qwen} 
\end{itemize}

We use the UMBRELA prompt \cite{Upadhyay2024umb} to obtain LLM relevance judgments across various LLM budgets $R$. For embeddings, we use the \texttt{all-MiniLM-L6-v2}~\cite{karpukhin2020dense} and the \texttt{msmarco-distilbert-base-tas-b} (~\citealp{hofstatter2021efficient}; see \autoref{sec:tasb}) We set the number of top passages to aggregate as \( T = 3 \) and define the aggregation function \( \phi \) (cf.~\Autoref{eq:item-level}) as the mean \(\mathrm{mean}(\cdot)\) over passage scores. The observation noise variance \(\alpha\) is set by default to \(10^{-3}\). Finally, the length scale \(\ell\) in the RBF kernel is optimized using the L-BFGS-B
~\cite{Zhu1997LBFGSB} algorithm.\footnote{Additional experiments analyzing the impact of different embeddings and hyperparameter choices are provided in \autoref{sec:tasb}, \Autoref{sec:GPR_hyper}, and \Autoref{sec:nlrec_hyper}} Evaluation is conducted using NDCG and Precision @10 and @30.

We address the following research questions:

\begin{description} 

\item[\textbf{RQ1 (Kernel Choice):}] Does the stationary RBF kernel outperform non-stationary kernels in relevance scoring?

\item[\textbf{RQ2 (Sampling Strategy):}] Does including a fraction \(\epsilon\) of randomly sampled exploratory passages with a cap \(\tau\) lead to better relevance scoring compared to only selecting top-ranked passages from DR?

\item[\textbf{RQ3 (Performance Comparison):}] Does GPR-LLM consistently outperform baseline methods across different LLM backbones given the same labeling budget?

\item[\textbf{RQ4 (Multimodal Relevance Scoring):}] Do empirical results support that GPR-LLM with the RBF kernel more effectively captures multimodal relevance compared to other methods?

\end{description}

\subsection{Results}

\begin{figure*}[h]
    \centering
    \includegraphics[width=\linewidth]{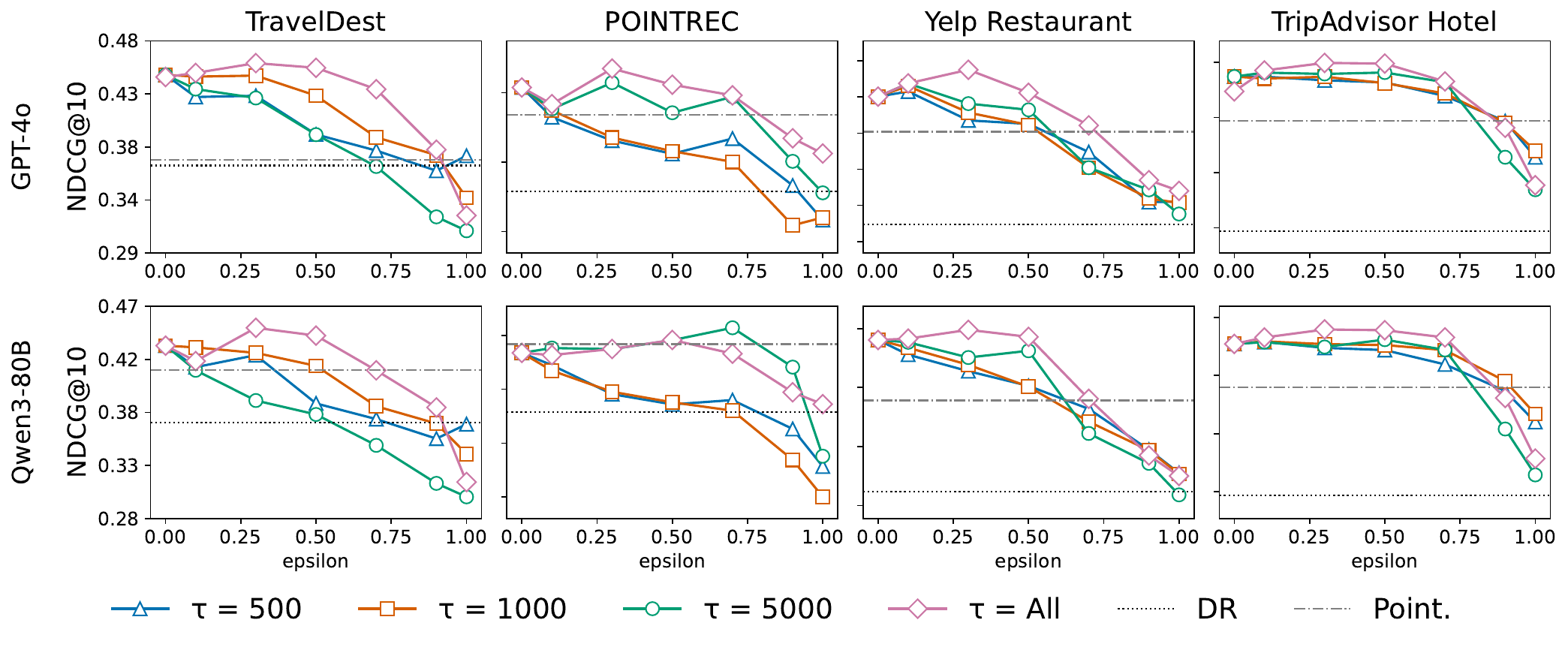}
    \caption{\textbf{RQ2:} Performance of GPR-LLM under $R=50$ with varying sampling sets. Each configuration labels an $\epsilon$ fraction of random passages from the top-$\tau$ non-DR set and a $(1-\epsilon)$ fraction from top DR-ranked passages.}
        \label{fig:rq2-e}
\end{figure*}

\begin{table*}[h!] \footnotesize
    \centering 
    \fitwide{
    \setlength{\tabcolsep}{3pt}
    \renewcommand{\arraystretch}{1.0} 
    \scriptsize
    \begin{tabular}{lllcccc|cccc|cccc|cccc}
    \toprule
    \multirow{2}{*}{\textbf{Budget}} & \multirow{2}{*}{\textbf{Backbone}} & \multirow{2}{*}{\textbf{Method}} &
    \multicolumn{4}{c|}{\textbf{TravelDest}} &
    \multicolumn{4}{c|}{\textbf{POINTREC}} &
    \multicolumn{4}{c|}{\textbf{Yelp Restaurant}} &
    \multicolumn{4}{c}{\textbf{TripAdvisor Hotel}}\\
    & & & P@10 & N@10 & P@30 & N@30 
            & P@10 & N@10 & P@30 & N@30 
            & P@10 & N@10 & P@30 & N@30 
            & P@10 & N@10 & P@30 & N@30\\
    \midrule
    \multirow{2}{*}{N/A} 
    & N/A & BM25 & 0.234 & 0.238 & 0.237 & 0.239
          & 0.025 & 0.032 & 0.025 & 0.038
          & 0.309 & 0.327 & 0.236 & 0.283
          & 0.205 & 0.257 & 0.153 & 0.325 \\
    & N/A & DR & 0.360 & 0.366 & 0.314 & 0.332
          & 0.164 & 0.179 & 0.104 & 0.182
          & 0.346 & 0.362 & 0.282 & 0.331
          & 0.231 & 0.297 & 0.166 & 0.365 \\
    \midrule
    \multirow{7}{*}{25}  
    & N/A & CE & 0.384 & 0.385 & 0.316 & 0.336
                     & 0.121 & 0.131 & 0.087 & 0.137
                     & 0.332 & 0.349 & 0.275 & 0.321
                     & 0.227 & 0.273 & 0.168 & 0.336 \\
    \cmidrule(lr){2-3}
    & \multirow{3}{*}{GPT-4o} & Point.   & 0.294 & 0.309 & 0.238 & 0.219
                     & \textbf{0.175} & \textbf{0.206} & 0.106 & 0.158
                     & 0.324 & 0.374 & 0.237 & 0.260
                     & 0.251 & 0.349 & 0.178 & 0.332 \\
    &  & GPR-LLM          & \textbf{0.376}$^*$ & \textbf{0.401}$^*$ & \textbf{0.340}$^*$ & \textbf{0.364}$^*$
                     & 0.157 & 0.200 & \textbf{0.106} & \textbf{0.158}
                     & \textbf{0.360} & \textbf{0.402}$^*$ & \textbf{0.273} & \textbf{0.340}$^*$
                     & \textbf{0.282}$^*$ & \textbf{0.382}$^*$ & \textbf{0.182} & \textbf{0.426}$^*$ \\
    \cmidrule(lr){2-3}
    & \multirow{3}{*}{Qwen3-80B} & Point. & 0.308 & 0.345 & 0.286 & 0.309
                     & 0.175 & 0.214 & \textbf{0.111} & \textbf{0.206}
                     & 0.336 & 0.381 & 0.266 & 0.331
                     & 0.258 & 0.353 & 0.165 & 0.398 \\
    &  & GPR-LLM & \textbf{0.366}$^*$ & \textbf{0.384}$^*$ & \textbf{0.285} & \textbf{0.316}
                     & \textbf{0.179} & \textbf{0.223} & 0.082 & 0.177
                     & \textbf{0.395}$^*$ & \textbf{0.437}$^*$ & \textbf{0.277} & \textbf{0.353}$^*$
                     & \textbf{0.270} & \textbf{0.365} & \textbf{0.176} & \textbf{0.418} \\
    \midrule
    \multirow{7}{*}{50}  
    & N/A & CE   & 0.390 & 0.399 & 0.325 & 0.348
                     & 0.093 & 0.096 & 0.077 & 0.110
                     & 0.314 & 0.334 & 0.272 & 0.315
                     & 0.216 & 0.258 & 0.167 & 0.329 \\
    \cmidrule(lr){2-3}
    & \multirow{3}{*}{GPT-4o} & Point.   & 0.356 & 0.371 & 0.248 & 0.281
                     & 0.196 & 0.234 & 0.105 & 0.208
                     & 0.386 & 0.426 & 0.254 & 0.332
                     & 0.289 & 0.397 & 0.169 & 0.417 \\
    &  & GPR-LLM          & \textbf{0.432}$^*$ & \textbf{0.445}$^*$ & \textbf{0.362}$^*$ & \textbf{0.389}$^*$
                     & \textbf{0.236}$^*$ & \textbf{0.262}$^*$ & \textbf{0.113} & \textbf{0.222}
                     & \textbf{0.408} & \textbf{0.451}$^*$ & \textbf{0.304}$^*$ & \textbf{0.377}$^*$
                     & \textbf{0.324}$^*$ & \textbf{0.439}$^*$ & \textbf{0.197}$^*$ & \textbf{0.482}$^*$ \\
    \cmidrule(lr){2-3}
    & \multirow{3}{*}{Qwen3-80B} & Point. & 0.392 & 0.415 & 0.273 & 0.315
                     & \textbf{0.211} & \textbf{0.242} & \textbf{0.108} & \textbf{0.213}
                     & 0.401 & 0.439 & 0.266 & 0.347
                     & 0.296 & 0.390 & 0.169 & 0.417 \\
    &  & GPR-LLM & \textbf{0.414}$^*$ & \textbf{0.437}$^*$ & \textbf{0.327}$^*$ & \textbf{0.363}$^*$
                     & 0.200 & 0.234 & 0.102 & 0.202
                     & \textbf{0.439}$^*$ & \textbf{0.490}$^*$ & \textbf{0.308}$^*$ & \textbf{0.398}$^*$
                     & \textbf{0.314} & \textbf{0.420}$^*$ & \textbf{0.196}$^*$ & \textbf{0.472}$^*$ \\
    \midrule
    \multirow{7}{*}{100}
    & N/A & CE   & 0.364 & 0.359 & 0.312 & 0.324
                     & 0.089 & 0.095 & 0.069 & 0.104
                     & 0.320 & 0.330 & 0.270 & 0.309
                     & 0.204 & 0.253 & 0.161 & 0.322 \\
    \cmidrule(lr){2-3}
    & \multirow{3}{*}{GPT-4o} & Point.   & 0.398 & 0.406 & 0.296 & 0.328
                     & 0.225 & 0.255 & 0.124 & 0.231
                     & 0.418 & 0.459 & 0.298 & 0.379
                     & 0.322 & 0.438 & 0.189 & 0.472 \\
    &  & GPR-LLM          & \textbf{0.448}$^*$ & \textbf{0.472}$^*$ & \textbf{0.380}$^*$ & \textbf{0.412}$^*$
                     & \textbf{0.246}$^*$ & \textbf{0.269} & \textbf{0.130} & \textbf{0.239}
                     & \textbf{0.444}$^*$ & \textbf{0.487}$^*$ & \textbf{0.334}$^*$ & \textbf{0.413}$^*$
                     & \textbf{0.358}$^*$ & \textbf{0.481}$^*$ & \textbf{0.218}$^*$ & \textbf{0.529}$^*$ \\
    \cmidrule(lr){2-3}
    & \multirow{3}{*}{Qwen3-80B} & Point. & 0.418 & 0.449 & 0.323 & 0.366
                     & 0.207 & 0.230 & \textbf{0.121} & \textbf{0.218}
                     & 0.424 & 0.457 & 0.303 & 0.381
                     & 0.319 & 0.425 & 0.192 & 0.467 \\
    &  & GPR-LLM & \textbf{0.472}$^*$ & \textbf{0.486}$^*$ & \textbf{0.362}$^*$ & \textbf{0.399}$^*$
                     & \textbf{0.225}$^*$ & \textbf{0.258}$^*$ & 0.123 & 0.226
                     & \textbf{0.491}$^*$ & \textbf{0.534}$^*$ & \textbf{0.355}$^*$ & \textbf{0.443}$^*$
                     & \textbf{0.345}$^*$ & \textbf{0.456}$^*$ & \textbf{0.220}$^*$ & \textbf{0.520}$^*$ \\
    \bottomrule
    \end{tabular}
    }
    \caption{\textbf{RQ3}: Comparison of DR, Cross-Encoder (CE), Pointwise LLM-based Relevance Scoring (Point.) and GPR-LLM (RBF kernel, $\epsilon = 0.1$, $\tau$ set to include all passages) across four datasets at varying LLM label budgets. 
    Metrics: Precision@10 (P@10), NDCG@10 (N@10), Precision@30 (P@30), NDCG@30 (N@30). 
    Bold values indicate the best-performing method between Point. and GPR-LLM for each backbone. Statistically significant improvements over Point. (paired \(t\)-test, \(p < 0.05\)) are indicated by an asterisk ($^*$).}
    \label{tab1:overall-perforamnce}
\end{table*}

\paragraph{RQ1 (Kernel Choice)}
To address \textbf{RQ1}, \Autoref{fig:kernel-comparison-rq1} compares different kernel functions (cf.~\Autoref{sec:kernel}) under greedy sampling ($\epsilon = 0$) with $\tau$ set to include all passages. The stationary RBF kernel consistently and significantly outperforms the non-stationary dot product and cosine similarity kernels across all datasets and LLM backbones. These results confirm the advantage of stationary kernel for GPR-LLM, likely due to its capability to capture the multimodal passage-level relevance scoring function. This is further examined in \textbf{RQ4}. Additionally, \Autoref{fig:kernel-comparison-rq1} demonstrates that GPR-LLM with the RBF kernel outperforms all baseline methods under the same labeling budget, even without tuning sampling parameters.


\paragraph{RQ2 (Sampling Strategy).} \Autoref{fig:rq2-e} shows the effect of different sampling configurations that mix a fraction $\epsilon$ of randomly selected passages from the top-$\tau$ non-DR set with $(1-\epsilon)$ passages drawn from the top DR rankings. Across all datasets and LLM backbones, a small exploratory fraction ($\epsilon = 0.3$) consistently improves GPR-LLM performance when the sampling range is sufficiently large ($\tau \geq 5000$), whereas no such gain is observed for smaller $\tau$. Under most sampling configurations, GPR-LLM also outperforms the pointwise LLM-based scoring baseline (dot–dash line). These results suggest that limited exploration enables GPR-LLM to better capture the multimodal relevance distribution by incorporating a more diverse subset of samples, but only when the exploration remains balanced and the sampled passages are sufficiently distinct from the top DR-ranking.

\paragraph{RQ3 (Performance Comparison)} \autoref{tab1:overall-perforamnce} compares GPR-LLM (with the best-performing RBF kernel, $\epsilon=0.1$, and $\tau$ including all passages) against all baseline methods across various LLM backbones and labeling budgets. GPR-LLM consistently outperforms baselines at same LLM budgets with only a few exceptions, and specifically achieves up to 65\% improvements over the pointwise LLM relevance scoring. Also, GPR-LLM is often able to outperform baselines that use twice as many labels. The performance is consistent across different NLRec datasets and LLM backbones, which highlights the robustness and effectiveness of our proposed method.

\begin{figure*}[htbp]
    \centering
    \includegraphics[width=\linewidth]{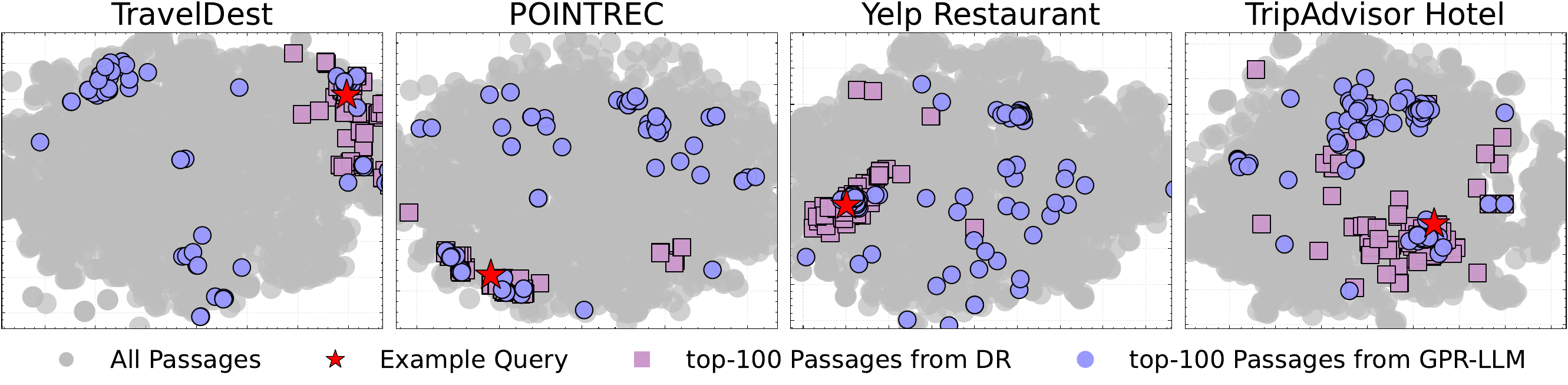}
  \caption{\textbf{RQ4:} Distribution of the top-100 passages from DR and GPR-LLM for example queries across all datasets. The embedding space is reduced to two dimensions using t-SNE for visualization. Passages ranked highest by DR (purple squares) cluster tightly around the query (red star), reflecting DR’s unimodal assumption. In contrast, GPR-LLM ({blue circles}) identifies passages dispersed across multiple regions, highlighting its ability to capture the multimodal relevance distribution hypothesized in~\autoref{fig:multimodal-dense}.}
        \label{fig:rq4}
\end{figure*}

\paragraph{RQ4 (Multimodal Relevance Scoring)}
Next, we investigate the underlying factors contributing to these performance gains. \Autoref{fig:rq4} visualizes the distribution of top-ranked passages for a representative query from each dataset using both DR and GPR-LLM in the reduced embedding space using t-SNE~\citep{vanDerMaaten2008tsne}.

DR’s top-ranked passages tend to cluster tightly around the query across all datasets, reflecting its unimodal relevance assumption with a single peak at the query (i.e., passages closer to the query are inherently more relevant). In contrast, GPR-LLM retrieves passages distributed across multiple distinct regions in the embedding space. This distribution aligns naturally with our multimodal relevance hypothesis illustrated in \Autoref{fig:multimodal-dense}, where each region represents a local relevance peak within the overall multimodal relevance surface. By capturing distant yet relevant passages, GPR-LLM overcomes a key limitation of DR. This multimodal retrieval pattern is consistently observed across all four benchmark datasets, providing a compelling explanation for GPR-LLM’s performance advantage. 
\section{Related Work}
\subsection{Natural Language Recommendation}

Natural Language Recommendation (NLRec) aims to generate item recommendations based on user-issued free-form textual requests \citep{Kang2017}. NLRec systems typically exhibit two key properties. First, they utilize natural language requests to encode user intent and preferences, contrasting with traditional recommender systems that mainly rely on structured interaction history (e.g., clicks, ratings) as their primary source of preference signals~\citep{ndr2017, bogers2018movie, bogers2019game, afzali2023pointrec}. Second, NLRec leverages collections of textual sources, such as item descriptions, reviews, or menus, to represent items and aggregate them into item-level representations~\citep{afzali2023pointrec, wen2025, recipe}.

As NLRec tasks grow in complexity, they often exhibit \textit{multimodal relevance}, characterized by multiple regions of high relevance scores in the dense embedding space (cf.~\autoref{fig:multimodal-dense}; \citealp{wen2025}). To effectively address this multimodal relevance, we propose GPR-LLM, which aims to model the multimodal relevance scoring function using Gaussian Process Regression with a stationary kernel.

\subsection{LLM Relevance Judgement}

Recently, Large Language Models (LLMs) have emerged as promising resources for determining the relevance between items or passages and natural language queries due to their strong contextual understanding and reasoning capabilities~\citep{LLM:qg, LLM:prp, LLM:rankgpt, LLM:zslistwise, LLM:yesno, LLM:zeroretriever}.

LLM-based relevance judgments generally fall into three categories: pointwise, listwise, and pairwise. Pointwise methods independently compute relevance scores for each query–passage pair~\citep{LLM:qg, LLM:yesno, Upadhyay2024umb}. Listwise methods prompt LLMs with a query and multiple candidate passages simultaneously, often using a sliding window approach to facilitate passage comparisons and directly generate a ranked list~\citep{LLM:rankgpt, LLM:zslistwise, LLM:zeroretriever}. Pairwise methods prompt LLMs to compare two passages, determine their relative relevance, and aggregate these pairwise preferences into final rankings~\citep{LLM:prp}.

To obtain passage-level relevance scores for aggregation into item-level scores in NLRec tasks, we primarily focus on pointwise LLM-based relevance scoring, which we compare against our proposed GPR-LLM method.

\section{Conclusion}

We introduced GPR-LLM, a method that models multimodal relevance scoring for NLRec using Gaussian Process Regression with a small subset of LLM-based relevance judgments. GPR-LLM addresses limitations of the unimodal scoring assumption in Dense Retrieval (DR) and reduces the reliance on exhaustive LLM labeling. Experiments across multiple datasets and LLM backbones demonstrate that GPR-LLM consistently outperforms baseline methods, including DR, Cross Encoder, and Pointwise LLM Relevance Scoring at same labeling budgets, and achieves comparable performance with substantially fewer LLM labels. These results establish GPR-LLM as both an efficient and effective approach for NLRec tasks.
\section*{Limitations}
While GPR-LLM achieves consistent improvement over baselines, several limitations remain. First, the performance of GPR-LLM critically depends on the quality and consistency of the LLM relevance judgments. Variations in LLM prompting or labeling criteria can impact accuracy and reliability, potentially influencing the quality of the GPR. 

Second, we use an $\epsilon$-greedy sampling strategy for exploration that depends on uniform random sampling. However, alternative sampling methods that explicitly leverage uncertainty-aware active learning or diversity-maximization exploration may align better with the exploration goal. Although such methods could further enhance performance, investigating them is beyond the scope of this paper and is thus left for future study. 

Third, we aggregate item-level scores using a fixed aggregation function $\phi$ (mean or max) over the top-$T$ passages. Alternative aggregation methods, such as harmonic mean or learned fusion networks, could be explored to potentially enhance performance. However, as our primary focus in this paper is on improving passage-level relevance scoring, we leave the investigation of more advanced aggregation methods for future work.

\section*{Ethical Considerations}
In deploying GPR-LLM, it is important to consider potential ethical implications. The reliance on LLMs introduces risks related to bias and fairness, as LLM relevance judgments may inherit or amplify biases present in training data. Thus, careful evaluation and monitoring of the generated labels and recommendation results are necessary to mitigate potential discriminatory impacts.

\bibliography{anthology-1,anthology-2,references}

\clearpage
\appendix

\section{UMBRELA Prompt for LLM Relevance Judgement}

We show the \texttt{UMBRELLA} prompt used to judge the relevance between a query and passages:

\begin{tcolorbox}[colback=gray!5!white, colframe=gray!75!black, 
title=UMBRELLA prompt, boxrule=0.3mm, width=0.5\textwidth, arc=3mm, auto outer arc=true]
Given a query and a list of passages, assign 
each passage a relevance score from 0 to 3:\newline
\newline
0 = Unrelated to the query.\newline
1 = Related but does not answer the query.\newline
2 = Partially answers the query but is \newline
unclear or mixed with extra information.\newline
3 = Fully dedicated to answering the query.\newline
\newline
Instructions:\newline
- Assign 1 if the passage is somewhat related 
but incomplete.\newline
- Assign 2 if it provides key information but 
includes unrelated content.\newline
- Assign 3 if it solely and fully addresses 
the query.\newline
- Otherwise, assign 0.\newline
\newline
Query: \texttt{\{query\}}\newline
Passages: \texttt{\{passages\}}\newline
\newline
Steps:\newline
1. Analyze the search intent.\newline
2. Measure content alignment (M).\newline
3. Assess passage trustworthiness (T).\newline
4. Decide on the final score (O).\newline
\newline
Return a JSON object mapping passage indices 
(starting from 0) to relevance scores.
Do not include any code in your response.\newline
\end{tcolorbox}

\section{Implementation Details}
\label{sec:impl_details}

We implement the following baseline methods for relevance scoring in Natural Language Recommendation (NLRec).

\paragraph{BM25.} 
We employ the Okapi BM25 algorithm~\cite{bm25} as implemented in the \texttt{Pyserini} toolkit, using default hyperparameters ($k_1 = 0.9$, $b = 0.4$).

\paragraph{Dense Retrieval (DR).} 
We use the \texttt{all-MiniLM-L6-v2} embedding model~\cite{karpukhin2020dense} with $D = 384$ and the \texttt{msmarco-distilbert-base-tas-b} model~\cite{hofstatter2021efficient} with $D = 768$ from Hugging Face. Candidate relevance scores are computed as the inner product between query and passage embeddings.

\paragraph{Cross-Encoder (CE).} 
We use the \texttt{cross-encoder/ms-marco-MiniLM-L-6-v2} model~\cite{reimers2019sentence,wolf2020transformers} from Hugging Face, which jointly encodes query–passage pairs and outputs a fine-grained scalar relevance score. The CE is applied to the top-$R$ passages retrieved by DR to maintain comparability with GPR-LLM and pointwise LLM-based scoring.

\paragraph{Pointwise LLM-based Relevance Scoring (Point.).} 
We use the commonly adopted UMBRELA \texttt{prompt}~\cite{Upadhyay2024umb} and follow the procedure of \citet{LLM:yesno} to obtain LLM-based passage relevance scores. For each query $q$ and passage $p_j$, the LLM outputs a vector of logits:
\begin{equation}
    \label{eq:llm_prompt}
    \mathbf{z} = \mathrm{LLM}(q, p_j, \texttt{prompt}) = [z_0, z_1, \dots, z_{K-1}],
\end{equation}
where each $z_k$ corresponds to the logit of a predefined discrete relevance label $r_k \in \{0, 1, \dots, K-1\}$.  
We compute the scalar LLM-based relevance score using the \textit{expected relevance} (ER) formulation:
\begin{equation}
    \label{eq:llm_er}
    \mathcal{S}_{i,j}^{\mathrm{LLM}} = \sum_{k=0}^{K-1} 
    \left( \frac{e^{z_k}}{\sum_{j=0}^{K-1} e^{z_j}} \right) \cdot r_k.
\end{equation}
We use both a closed-source LLM (\texttt{GPT-4o}; \citealp{openai2023gpt4}) and an open-source LLM (\texttt{Qwen-Next-80B}; \citealp{imple:qwen}) for generating relevance judgments. Pointwise LLM-based scoring is applied to the top-$R$ passages retrieved by DR to ensure fair comparability with GPR-LLM.

\section{How Sampling Quality Affects GPR-LLM}
\label{sec:subset_comp}

To better understand how the quality of the small labeled subset of passages influences model performance, we conduct a controlled experiment to examine whether augmenting the sampled subset with additional high-relevance passages can further improve GPR-LLM and how different kernel choices respond to such changes. The goal is to test a simple hypothesis: \emph{if the sampling process were improved, could GPR-LLM achieve higher performance?}

\paragraph{Experimental Setup.}
We use the \texttt{TravelDest} dataset and begin with a baseline of 100 passages selected greedily by dense-retrieval scores. We then incrementally augment this subset by adding 1–20 additional passages drawn from two distinct supervision pools:
\begin{enumerate}
    \item \textbf{High-relevance augmentation:} passages assigned high LLM relevance scores (2–3);
    \item \textbf{Low-relevance augmentation:} passages assigned low or partial relevance scores (1–2).
\end{enumerate}

\begin{figure}[t]
    \centering
    \includegraphics[width=1\linewidth]{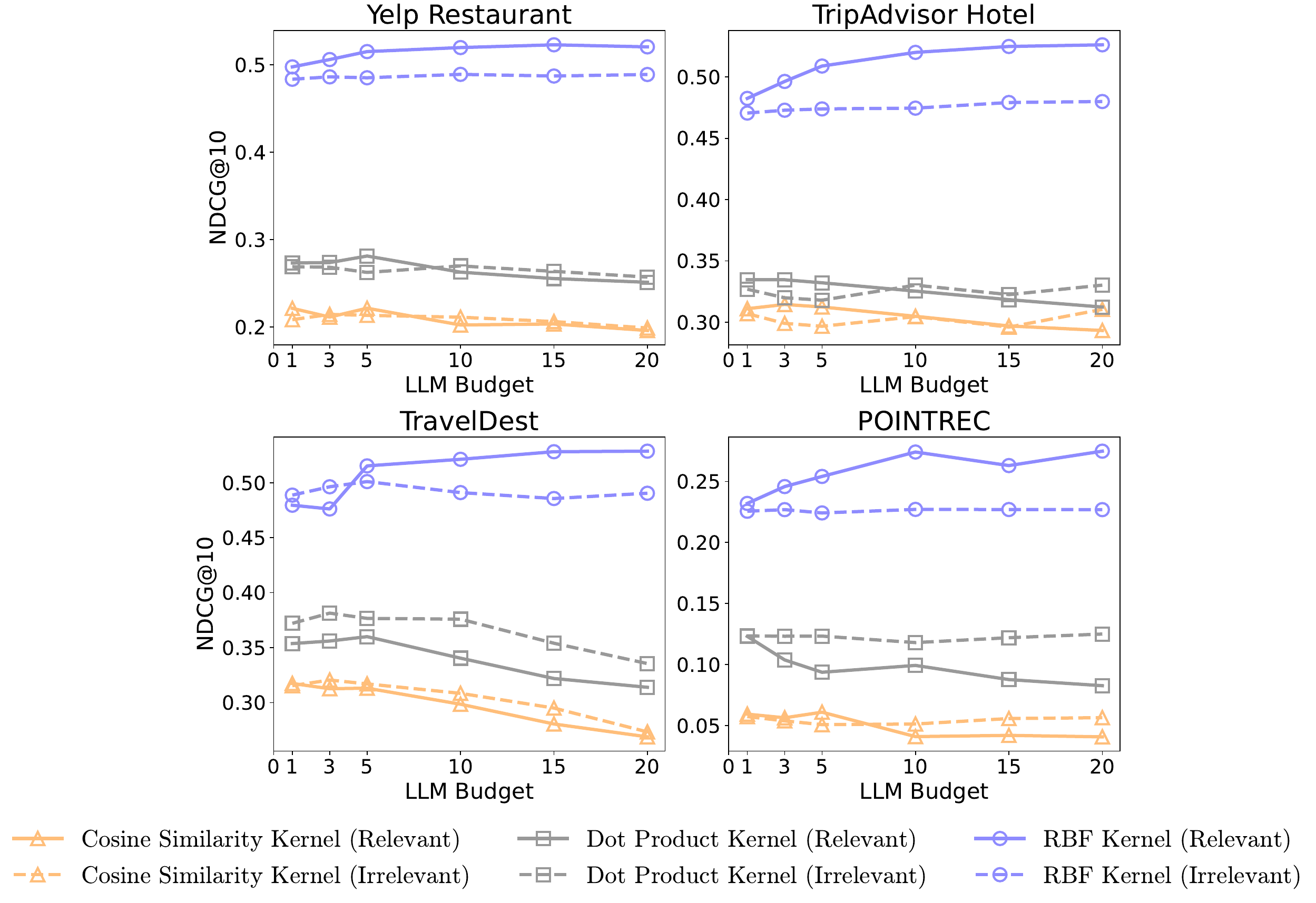}
    \caption{Effect of adding high-relevance vs.\ low-relevance passages to the sampling subset on \texttt{TravelDest}. We report NDCG@10 as additional passages (1–20) are added. Only the RBF kernel shows consistent gains when augmented with high-relevance labels, while dot product and cosine kernels remain insensitive to either augmentation type.}
    \label{fig:rel-irel}
\end{figure}

\paragraph{Results.}
Results are shown in \Autoref{fig:rel-irel}. Adding low-relevance passages yields negligible or inconsistent changes across all kernels, suggesting that weakly informative samples provide limited benefit. In contrast, augmenting with high-relevance passages leads to clear gains \emph{only} for the RBF kernel, which consistently improves as more high-quality samples are included. The stationary dot product and cosine kernels remain largely unaffected.

These results indicate that the stationary RBF kernel can better capture multimodal relevance scoring functions when provided with higher-quality labeled samples. This finding motivates us to move beyond purely greedy sampling based on DR rankings toward the proposed $\epsilon$-greedy strategy, which introduces exploration to identify additional relevant passages. It also suggests promising future directions such as \emph{uncertainty-aware active sampling} (e.g., GP-UCB~\cite{srinivas2009gaussian}) and \emph{diversity-optimized sampling} (e.g., core-set selection~\cite{sener2017active}), which could further enhance sampling efficiency and model performance.

\section{Impact of Different Embeddings}
\label{sec:tasb}

\begin{table*}[htbp]
\centering
\fitwide{
\setlength{\tabcolsep}{3pt}
\renewcommand{\arraystretch}{1.0}
\scriptsize
\begin{tabular}{llcccc|cccc|cccc|cccc}
\toprule
\multirow{2}{*}{\textbf{Budget}} & \multirow{2}{*}{\textbf{Method}} &
\multicolumn{4}{c|}{\textbf{TravelDest}} &
\multicolumn{4}{c|}{\textbf{POINTREC}} &
\multicolumn{4}{c|}{\textbf{Yelp Restaurant}} &
\multicolumn{4}{c}{\textbf{TripAdvisor Hotel}} \\
& & P@10 & N@10 & P@30 & N@30 
  & P@10 & N@10 & P@30 & N@30 
  & P@10 & N@10 & P@30 & N@30 
  & P@10 & N@10 & P@30 & N@30 \\
\midrule
\multirow{2}{*}{N/A} 
& DPR & 0.358 & 0.365 & 0.318 & 0.333
      & 0.143 & 0.161 & 0.094 & 0.165
      & 0.363 & 0.385 & 0.294 & 0.351
      & 0.224 & 0.275 & 0.165 & 0.349  \\
& BM25 & 0.234 & 0.238 & 0.237 & 0.239
      & 0.025 & 0.032 & 0.025 & 0.038
      & 0.309 & 0.327 & 0.236 & 0.283
      & 0.205 & 0.257 & 0.153 & 0.325 \\
\midrule
\multirow{2}{*}{25}  
& Point. & 0.344 & 0.379 & 0.313 & 0.339
              & 0.154 & 0.188 & 0.094 & 0.178
              & 0.340 & 0.382 & \textbf{0.289} & \textbf{0.355}
              & 0.227 & 0.302 & 0.165 & 0.369 \\
& GPR-LLM & \textbf{0.370} & \textbf{0.402} & \textbf{0.325} & \textbf{0.356}
              & \textbf{0.159} & \textbf{0.192} & \textbf{0.098} & \textbf{0.183}
              & \textbf{0.359} & \textbf{0.397} & 0.272 & 0.343
              & \textbf{\underline{0.286}} & \textbf{\underline{0.378}} & \textbf{0.177} & \textbf{\underline{0.419}} \\
\midrule
\multirow{2}{*}{50}  
& Point. & 0.380 & 0.419 & 0.298 & 0.339
              & 0.171 & 0.215 & 0.098 & 0.193
              & 0.368 & 0.410 & 0.282 & 0.355
              & 0.251 & 0.336 & 0.166 & 0.389 \\
& GPR-LLM & \textbf{\underline{0.416}} & \textbf{\underline{0.453}} & \textbf{\underline{0.348}} & \textbf{\underline{0.386}}
              & \textbf{0.176} & \textbf{0.217} & \textbf{0.103} & \textbf{0.197}
              & \textbf{0.376} & \textbf{0.414} & \textbf{0.286} & \textbf{0.361}
              & \textbf{0.325} & \textbf{0.199} & \textbf{0.470} \\
\midrule
\multirow{2}{*}{100}  
& Point. & 0.428 & 0.467 & 0.323 & 0.371
              & 0.214 & 0.245 & 0.102 & 0.200
              & 0.391 & 0.432 & 0.285 & 0.366
              & 0.273 & 0.366 & 0.170 & 0.408 \\
& GPR-LLM & \textbf{0.444} & \textbf{0.479} & \textbf{0.344} & \textbf{0.389}
              & \textbf{0.215} & \textbf{0.246} & \textbf{0.121} & \textbf{0.227}
              & \textbf{0.430} & \textbf{0.472} & \textbf{0.311} & \textbf{0.398}
              & \textbf{0.360} & \textbf{0.476} & \textbf{0.225} & \textbf{0.535} \\
\bottomrule
\end{tabular}
}
\caption{
Comparison of BM25, DPR, Pointwise LLM-based Scoring (Point.), and GPR-LLM (RBF kernel, $\epsilon = 0.1$, $\tau$ includes all passages) using the \texttt{msmarco-distilbert-base-tas-b} encoder with GPT-4o as the LLM backbone across four datasets: \texttt{TravelDest}, \texttt{POINTREC}, \texttt{Yelp Restaurant}, and \texttt{TripAdvisor Hotel}, under varying LLM labeling budgets. Metrics reported are Precision@10 (P@10), NDCG@10 (N@10), Precision@30 (P@30), and NDCG@30 (N@30). Bold values indicate the best performance in each column.
}
\label{tab:tasb}
\end{table*}

\Autoref{tab:tasb} presents the results obtained using the \texttt{msmarco-distilbert-base-tas-b} encoder with GPT-4o as the LLM backbone. The results follow the same overall trend observed with the \texttt{MiniLM-L6-v2} encoder, where GPR-LLM consistently outperforms both pointwise LLM-based scoring and other baseline methods across all datasets and LLM labeling budgets. The performance gains are most pronounced at smaller labeling budgets, indicating that GPR-LLM effectively leverages limited high-quality supervision to model the underlying multimodal relevance function. This consistent pattern across distinct embedding models demonstrates that the improvements of GPR-LLM stem from its core modeling design rather than encoder-specific characteristics.

\section{Hyperparameter Settings for GPR}
\label{sec:GPR_hyper}

\subsection{Observation Noise Variance $\alpha$.}  
\begin{figure}[H]
    \centering
    \includegraphics[width=1.0\linewidth]{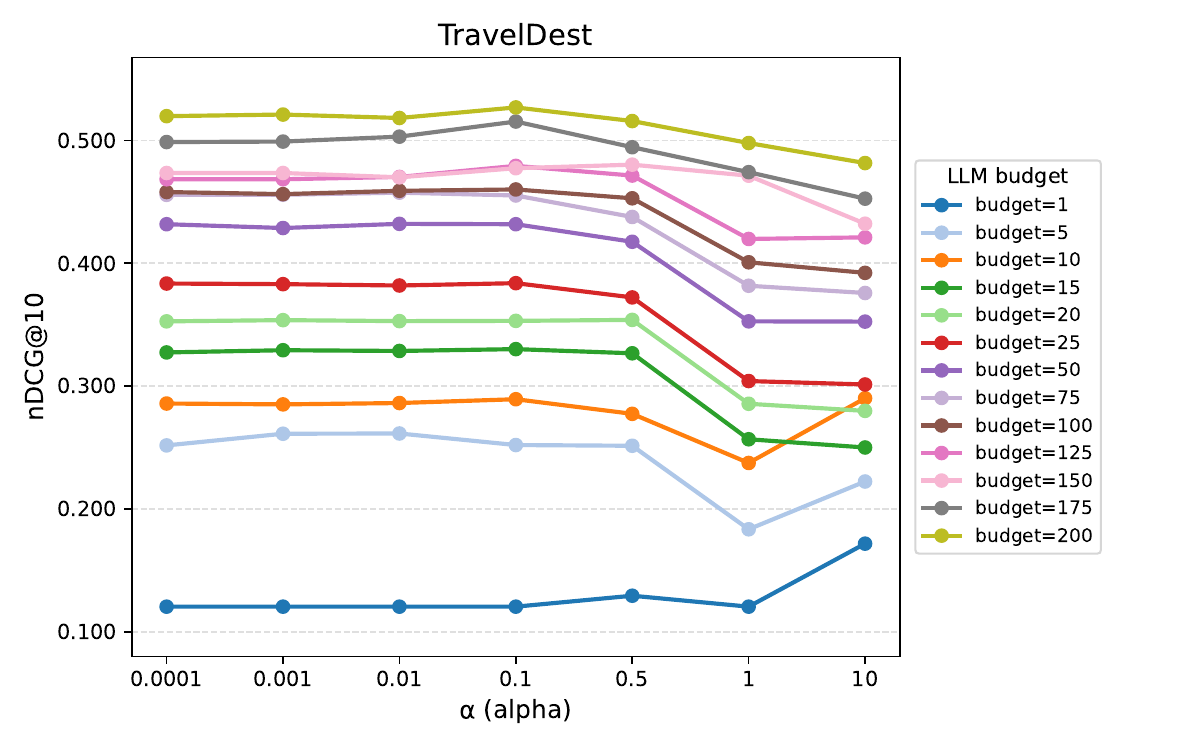}
    \caption{Performance of GPR-LLM in TravelDest with different values of  $\alpha$ of GPR using LLM budget of 1 to 200 labels.}
        \label{fig:travel_dest_ndcg_alpha}
\end{figure}

\begin{figure}[H]
    \centering
    \includegraphics[width=1.0\linewidth]{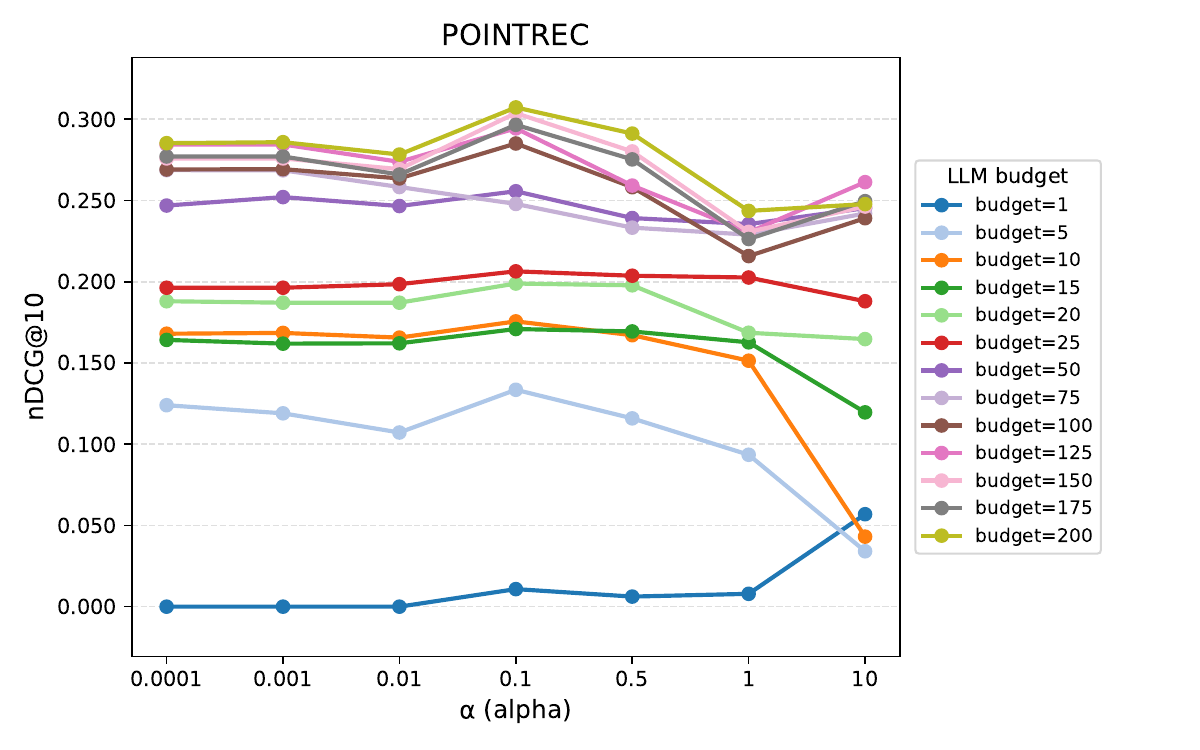}
    \caption{Performance of GPR-LLM in POINTREC with different values of  $\alpha$ of GPR using LLM budget of 1 to 200 labels.}
        \label{fig:pointrec_ndcg_alpha}
\end{figure}

\begin{figure}[H]
    \centering
    \includegraphics[width=1.0\linewidth]{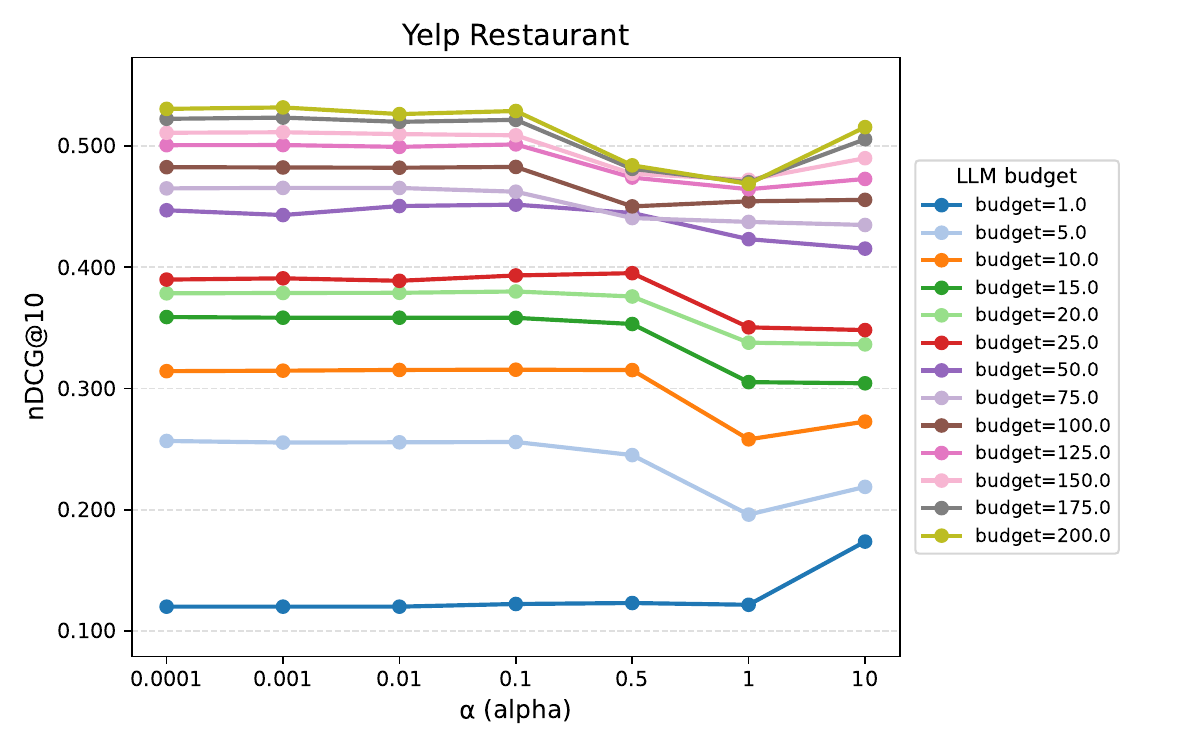}
    \caption{Performance of GPR-LLM in Yelp Restaurant with different values of  $\alpha$ of GPR using LLM budget of 1 to 200 labels.}
        \label{fig:restaurant_ndcg_alpha}
\end{figure}

\begin{figure}[H]
    \centering
    \includegraphics[width=1.0\linewidth]{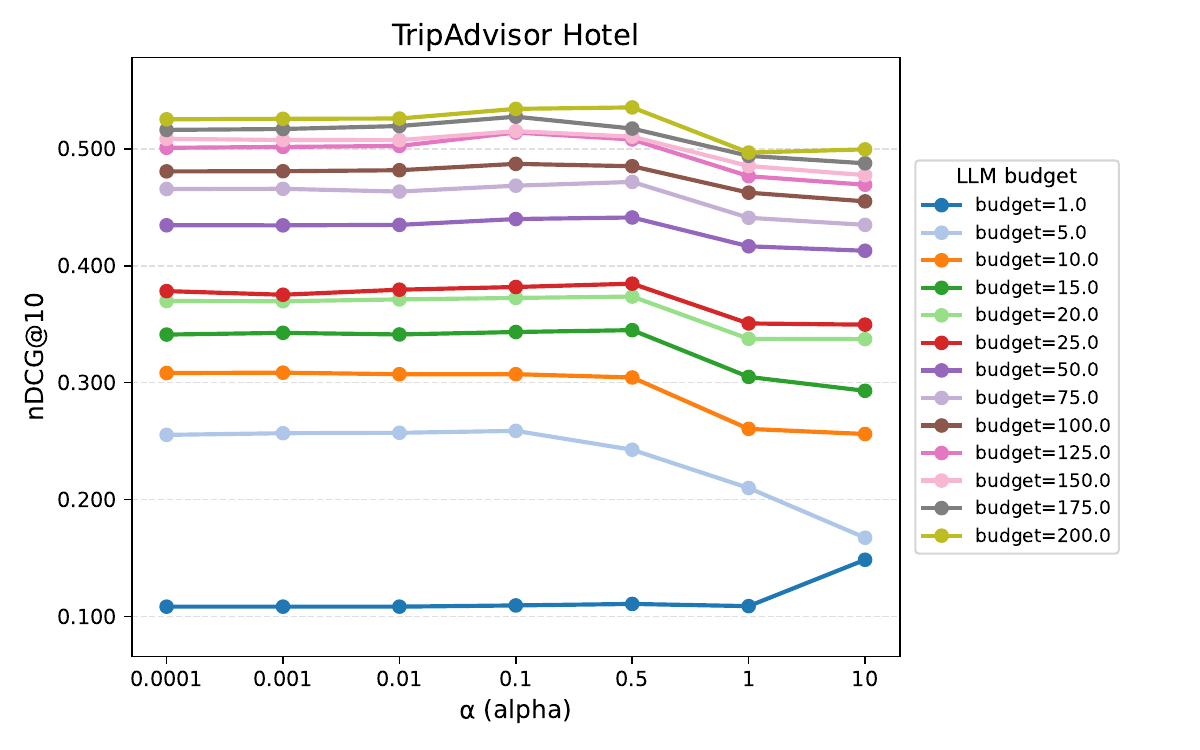}
    \caption{Performance of GPR-LLM in  TripAdvisor Hotel with different values of  $\alpha$ of GPR using LLM budget of 1 to 200 labels.}
        \label{fig:hotel_ndcg_alpha}
\end{figure}

In a noisy observation setting (i.e., the ground truth relevance is unknown), the performance of GPR can be affected by the variance of the Gaussian observation noises $\alpha$. Thus, we examine the difference in GPR-LLM performance under various values of $\alpha$. In \autoref{fig:travel_dest_ndcg_alpha}, \autoref{fig:pointrec_ndcg_alpha}, \autoref{fig:restaurant_ndcg_alpha} and \autoref{fig:hotel_ndcg_alpha}, we presents the performance of GPR-LLM under nDCG@10 using different value of $\alpha$. 

Overall, we observe a non-monotonic relationship between $\alpha$ and performance. Extremely low noise levels (e.g., $\alpha = 10^{-4}$, effectively assuming near-perfect LLM relevance judgments) tend to \emph{underperform}, likely due to overfitting to the few LLM judgments. Conversely, very high noise settings (e.g., $\alpha = 10$, treating LLM relevance judgments as extremely noisy) also degrade performance by underfitting the relevance signal. 

In all cases, moderate noise variance yields the best results: performance generally peaks at intermediate $\alpha$ values (around $10^{-2}$ to $10^{-1}$ in our experiments) and remains fairly stable across a broad mid-range. This trend is consistent across datasets, though the optimal $\alpha$ can vary slightly by dataset (each dataset’s curve achieves its maximum nDCG at a slightly different $\alpha$). Crucially, adding a reasonable amount of observation noise improves generalization, but too much noise or none at all is detrimental.

\subsection{Length Scale of the RBF Kernel.}  
The RBF kernel in GPR is defined as
\begin{equation}
    k(\mathbf{v}, \mathbf{v}') = \exp\left(-\frac{\|\mathbf{v} - \mathbf{v}'\|^2}{2\ell^2}\right)
\end{equation}
where \( \ell \) is the length scale hyperparameter that controls the smoothness of the learned function. Smaller values of \( \ell \) yield more flexible, highly localized functions, while larger values impose smoother, more global behavior. 

\begin{figure}[H]
    \centering
    \includegraphics[width=1.0\linewidth]{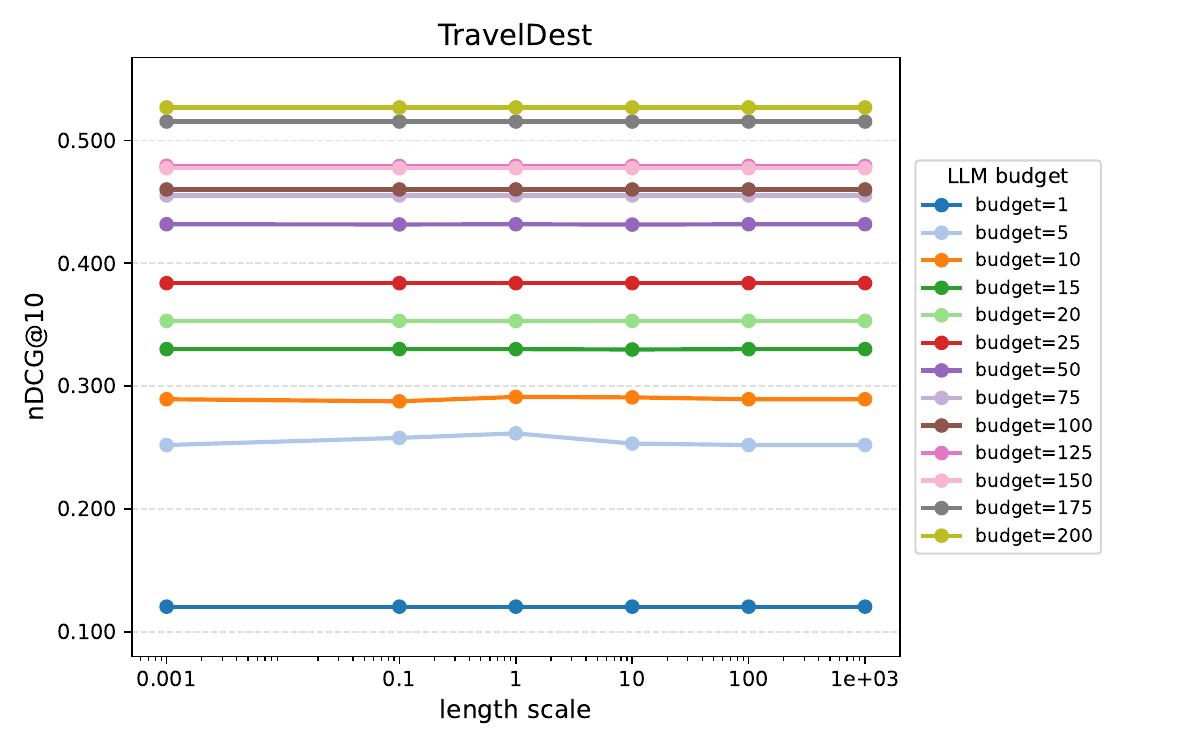}
    \caption{Performance of GPR-LLM in TravelDest across different value of\( \ell \) using LLM budget of 1 to 200 labels with L-BFGS-B algorithm.}
        \label{fig:travel_dest_ndcg_ls}
\end{figure}

\begin{figure}[H]
    \centering
    \includegraphics[width=1.0\linewidth]{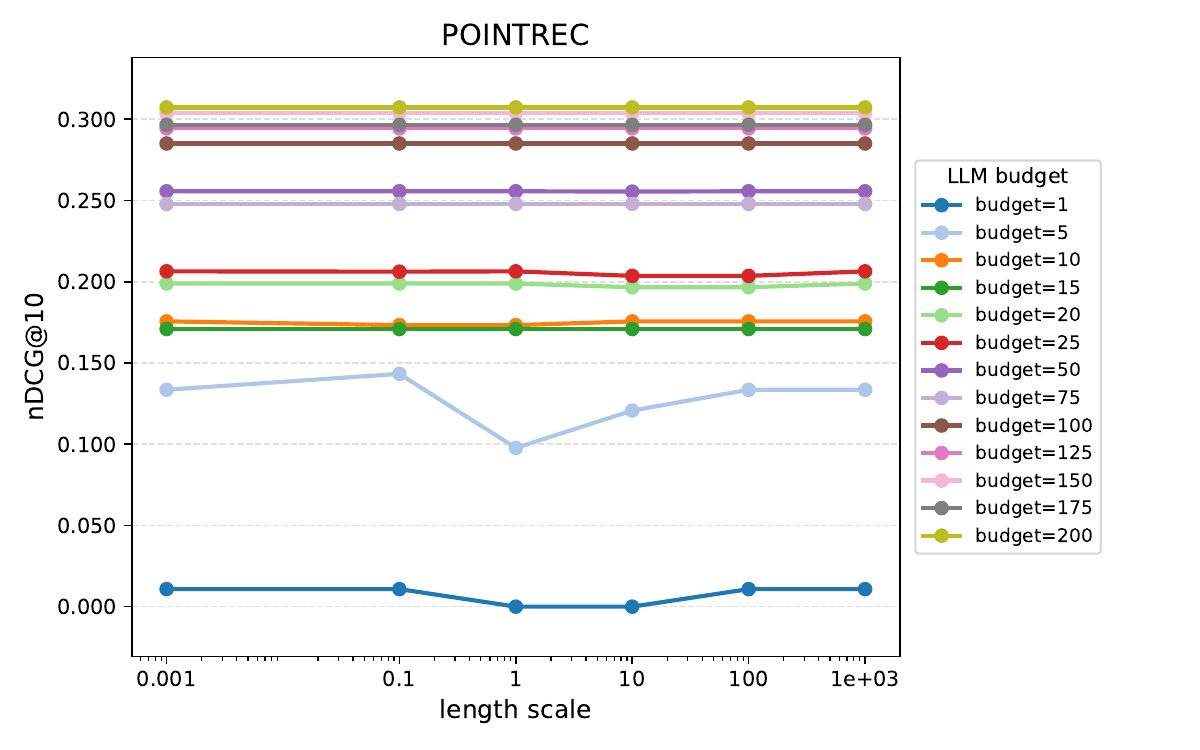}
    \caption{Performance of GPR-LLM in  POINTREC across different values of \( \ell \) using LLM budget of 1 to 200 labels with L-BFGS-B algorithm.}
        \label{fig:pointrec_ndcg_ls}
\end{figure}

\begin{figure}[H]
    \centering
    \includegraphics[width=1.0\linewidth]{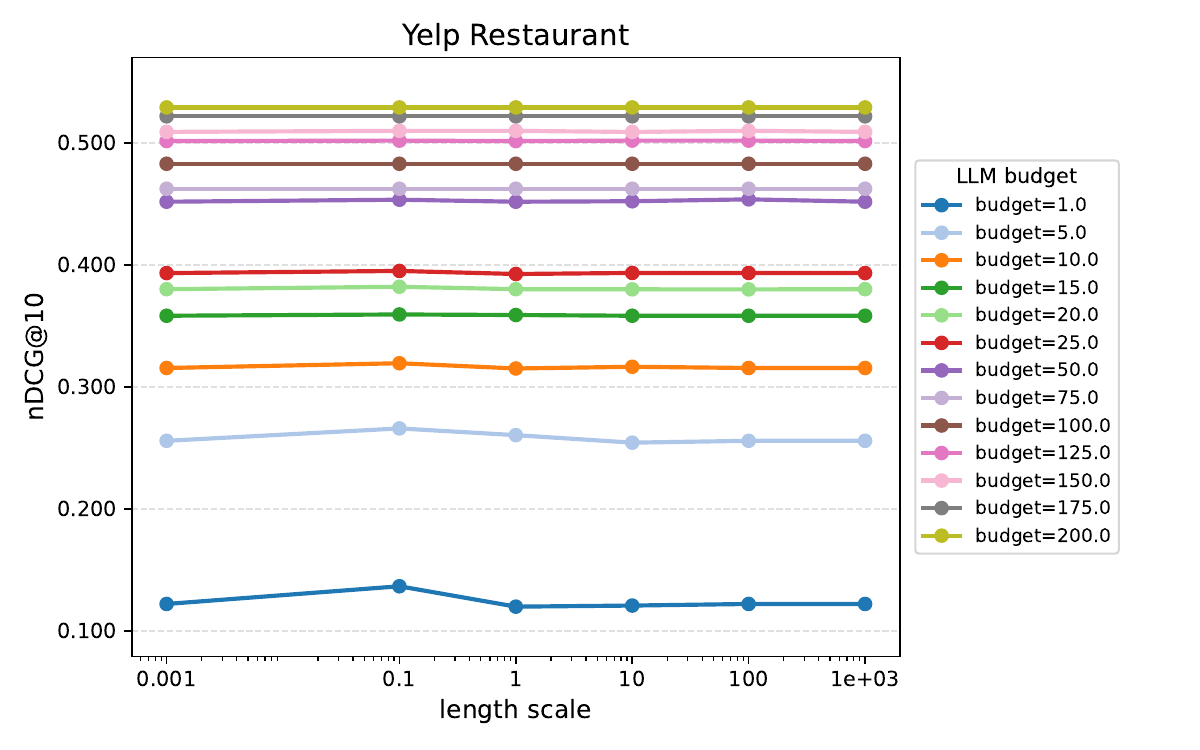}
    \caption{Performance of GPR-LLM in  Yelp Restaurant across different values of \( \ell \) using LLM budget of 1 to 200 labels with L-BFGS-B algorithm.}
        \label{fig:restaurant_ndcg_ls}
\end{figure}

\begin{figure}[H]
    \centering
    \includegraphics[width=1.0\linewidth]{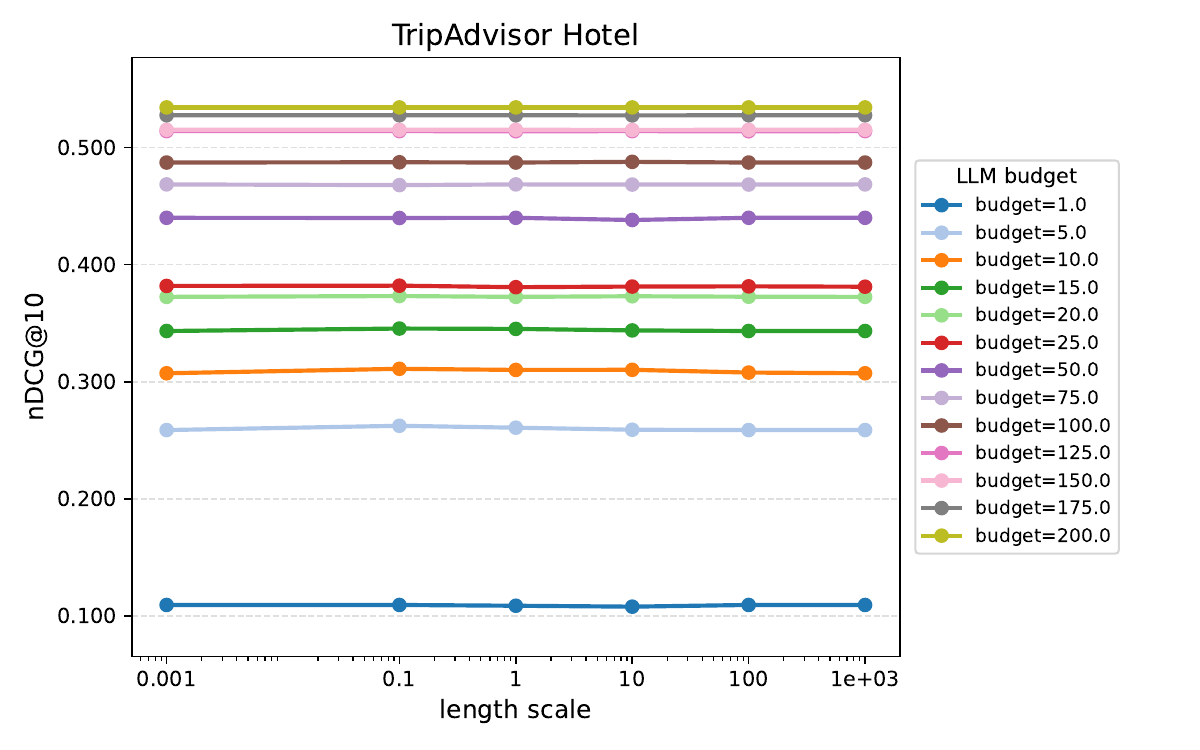}
    \caption{Performance of GPR-LLM in  TripAdvisor Hotel across different values of \( \ell \) using LLM budget of 1 to 200 labels with L-BFGS-B algorithm.}
        \label{fig:hotel_ndcg_ls}
\end{figure}

While \( \ell \) is optimized using L-BFGS-B algorithm, we empirically evaluate the impact of varying the initial value of  \( \ell \) over the range \([0.001,\, 0.1,\, 1,\, 10,\, 100,\, 1000]\). As shown in \autoref{fig:travel_dest_ndcg_ls}, \autoref{fig:pointrec_ndcg_ls},
\autoref{fig:restaurant_ndcg_ls} and \autoref{fig:hotel_ndcg_ls}, we can conclude that our algorithm is robust to different initializations of length scale values except for very small budget of LLM labels, where noise could impact the results.

\begin{figure}[H]
    \centering
    \includegraphics[width=1.0\linewidth]{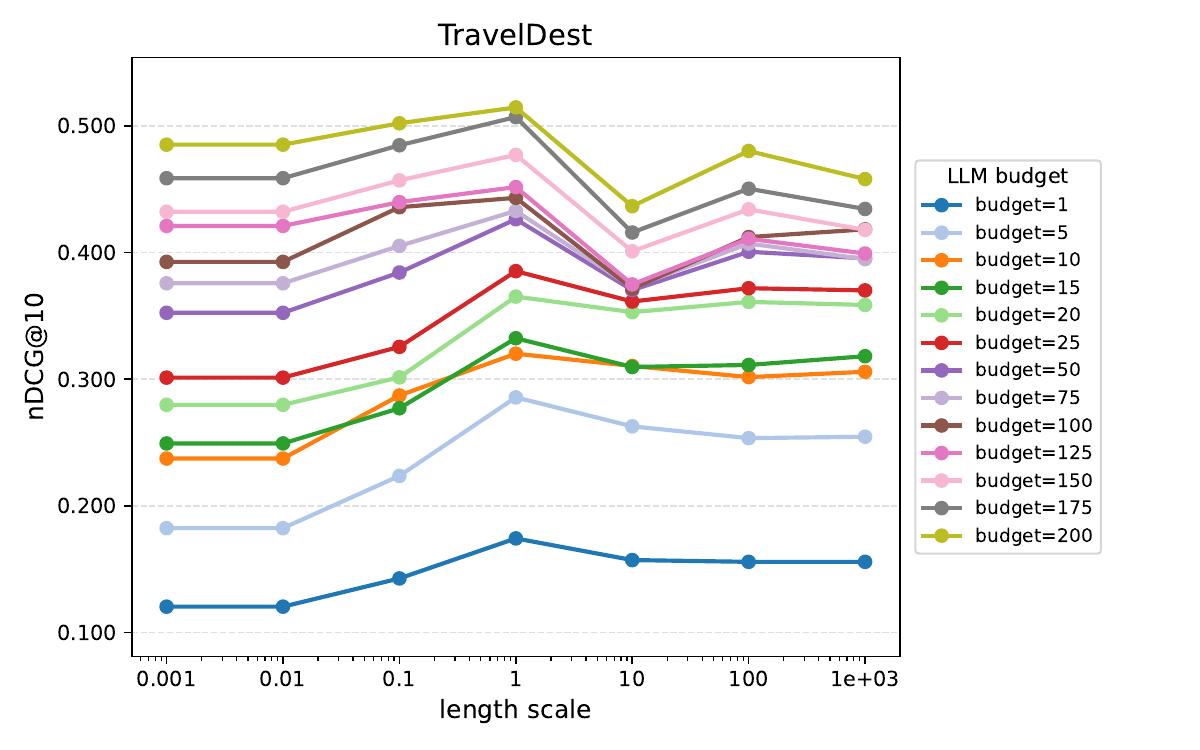}
    \caption{Performance of GPR-LLM in TravelDest across different values of \( \ell \) using LLM budget of 1 to 200 labels without L-BFGS-B algorithm.}
        \label{fig:travel_dest_ndcg_ls_new}
\end{figure}

\begin{figure}[H]
    \centering
    \includegraphics[width=1.0\linewidth]{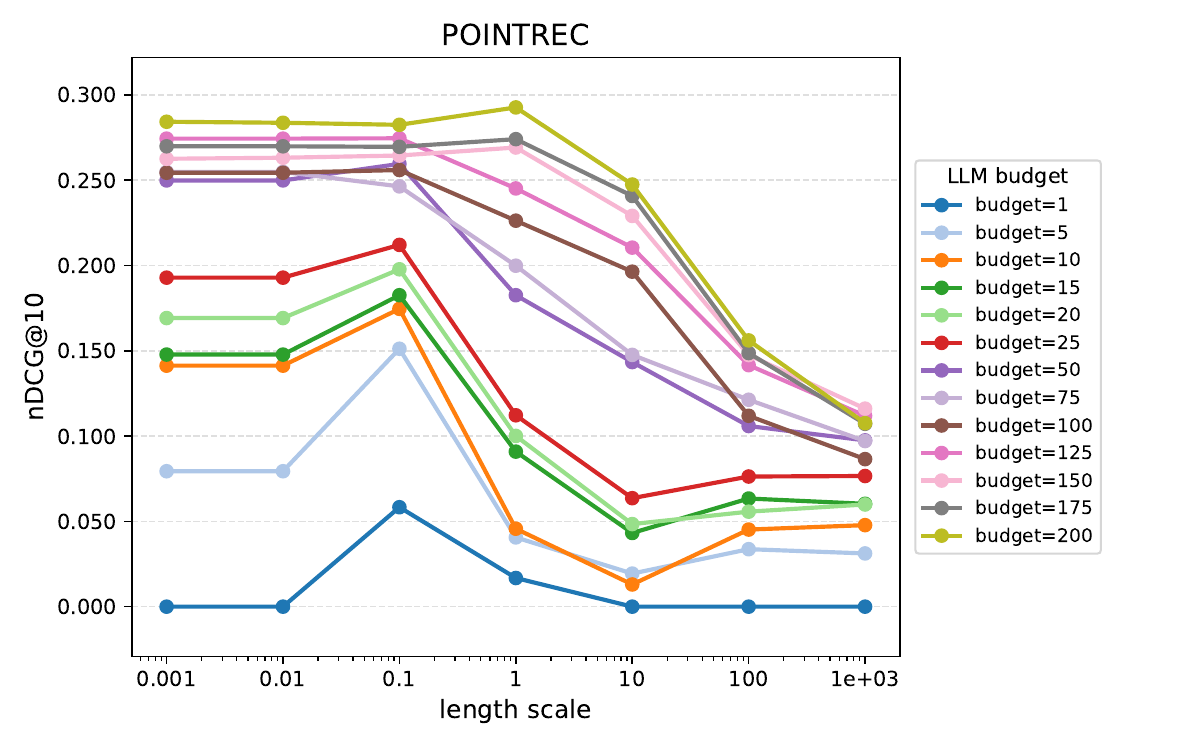}
    \caption{Performance of GPR-LLM in POINTREC across different values of \( \ell \) using LLM budget of 1 to 200 labels without L-BFGS-B algorithm.}
        \label{fig:pointrec_ndcg_ls_new}
\end{figure}

\begin{figure}[H]
    \centering
    \includegraphics[width=1.0\linewidth]{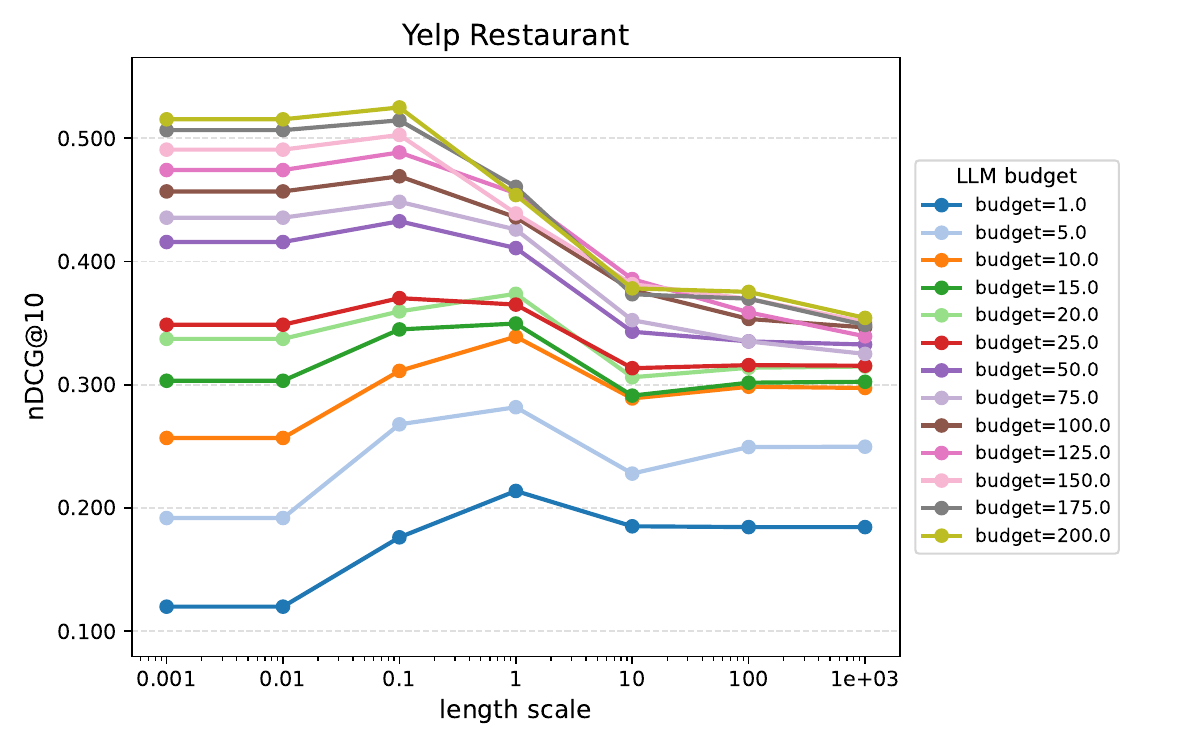}
    \caption{Performance of GPR-LLM in Yelp Restaurant across different values of \( \ell \) using LLM budget of 1 to 200 labels without L-BFGS-B algorithm.}
        \label{fig:restaurant_ndcg_ls_new}
\end{figure}

\begin{figure}[H]
    \centering
    \includegraphics[width=1.0\linewidth]{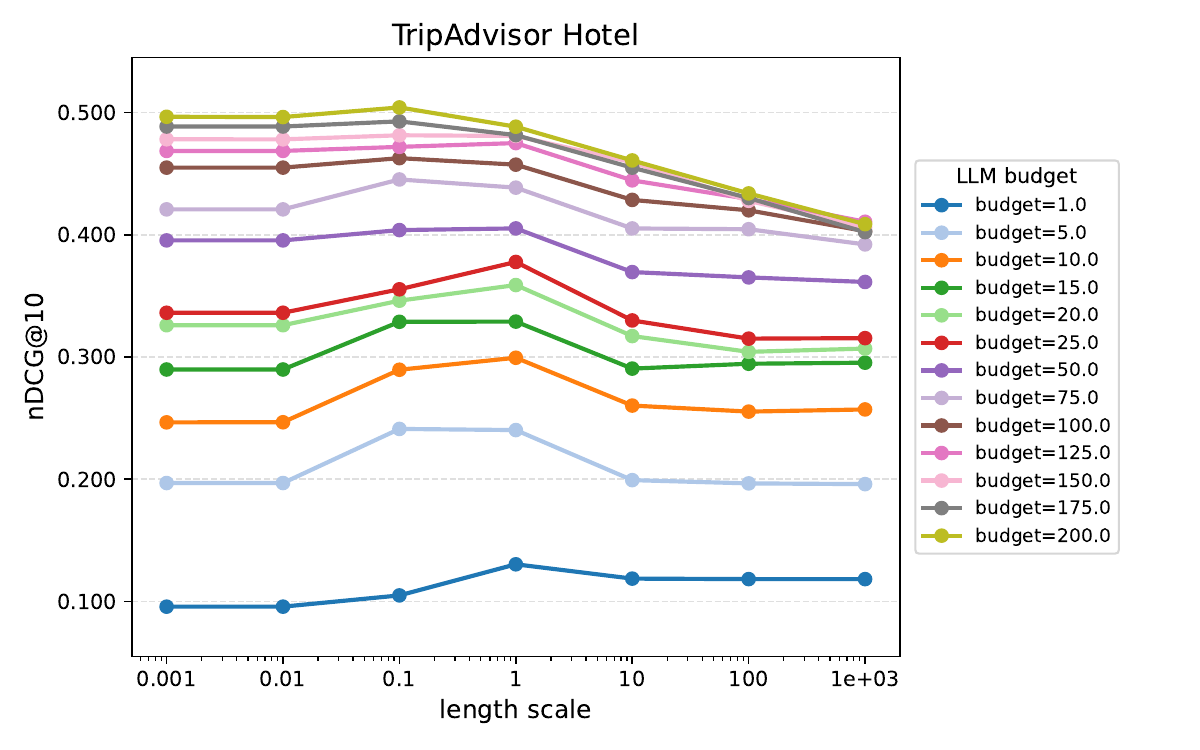}
    \caption{Performance of GPR-LLM in  TripAdvisor Hotel across different value of\( \ell \) using LLM budget of 1 to 200 labels without L-BFGS-B algorithm.}
        \label{fig:hotel_ndcg_ls_new}
\end{figure}

We also evaluate performance across different values of \( \ell \) without applying the L-BFGS-B algorithm. As shown in \autoref{fig:travel_dest_ndcg_ls_new}, \autoref{fig:pointrec_ndcg_ls_new}, \autoref{fig:restaurant_ndcg_ls_new}, and \autoref{fig:hotel_ndcg_ls_new}, under a fixed \( \alpha \), the best performance achieved without L-BFGS-B is comparable to that with L-BFGS-B. This suggests that L-BFGS-B provides an efficient and effective method for kernel parameter selection in GP-LLM.

\subsection{Scaling of Relevance Scores.} 
\begin{figure}[H]
    \centering
    \includegraphics[width=1.0\linewidth]{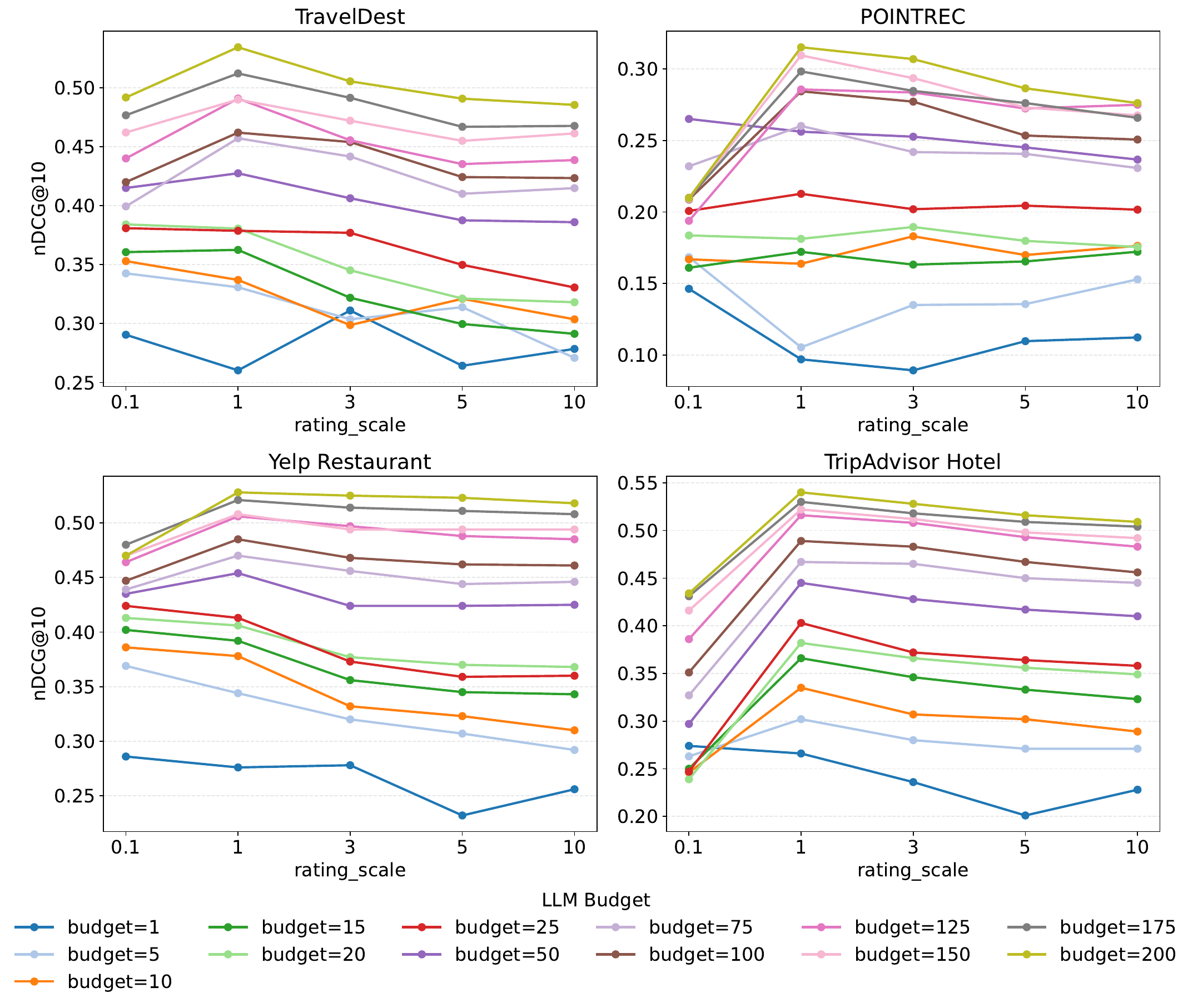}
    \caption{TravelDest, POINTREC, Yelp Restaurant, and TripAdvisor Hotel under various rating scale using LLM budget of 1 to 200 labels without L-BFGS-B algorithm.}
        \label{fig:rating_scale_combined}
\end{figure}

Figure~\ref{fig:rating_scale_combined} illustrates the effect of varying the rating scale used by the LLM to provide relevance judgments, with results shown for all four datasets (nDCG@10 across different LLM budgets of labels). Across all four datasets, applying extreme scaling to the LLM relevance scores consistently degrades GP-LLM’s nDCG@10 performance, whereas keeping the labels at or near their original scale yields the best results. 

In particular, using the original rating range (scale factor = 1) produces the highest nDCG@10 on TravelDest, POINTREC, Yelp Restaurant, and TripAdvisor Hotel, for both low and high LLM budgets. Any substantial compression of the label range (e.g., a 0.1× factor) or expansion (e.g., a 10× factor) leads to a notable drop in ranking effectiveness across all LLM budgets. 

This suggests that over-compressing the relevance scores blurs meaningful differences between items, while over-expanding them amplifies noise and overemphasizes minor relevance distinctions, in both cases hurting the GP-LLM’s ability to accurately rank items. Consequently, maintaining the original scale preserves the proper balance of signal to noise in the LLM labels and achieves the strongest overall nDCG@10 performance in the GP-LLM framework.

\section{Hyperparameter Settings for Item-Level Relevance Aggregation}
\label{sec:nlrec_hyper}

The final item-level score \( \mathcal{S}_i \) for item \( i \in \mathcal{I} \) is computed by aggregating the top-\(T\) passage-level scores:
\begin{equation}
    \mathcal{S}_i = \phi\left(\left\{\, \hat{f}_q(\mathbf{v}_{p_j}) \mid p_j \in \operatorname{top}_T(\mathcal{P}_i, \hat{f}_q) \,\right\}\right),
\end{equation}
where \( \phi \) is the aggregation function (e.g., mean, max) and \( \operatorname{top}_T(\mathcal{P}_i, \hat{f}_q) \) returns the top-\(T\) passages for item \( i \) ranked by the GPR relevance scoring function \( \hat{f}_q \).

We vary the number \( T \) of top-ranked passages used in aggregation and evaluate its effect on item-level ranking performance. A small \( T \) may ignore informative passages, while a large \( T \) may include irrelevant noise. We vary it by [1, 3, 5, 10, 25, 50] and set LLM budget and epsilon to 100 and 0 respectively. 

\begin{figure}[H]
    \centering
    \includegraphics[width=1.0\linewidth]{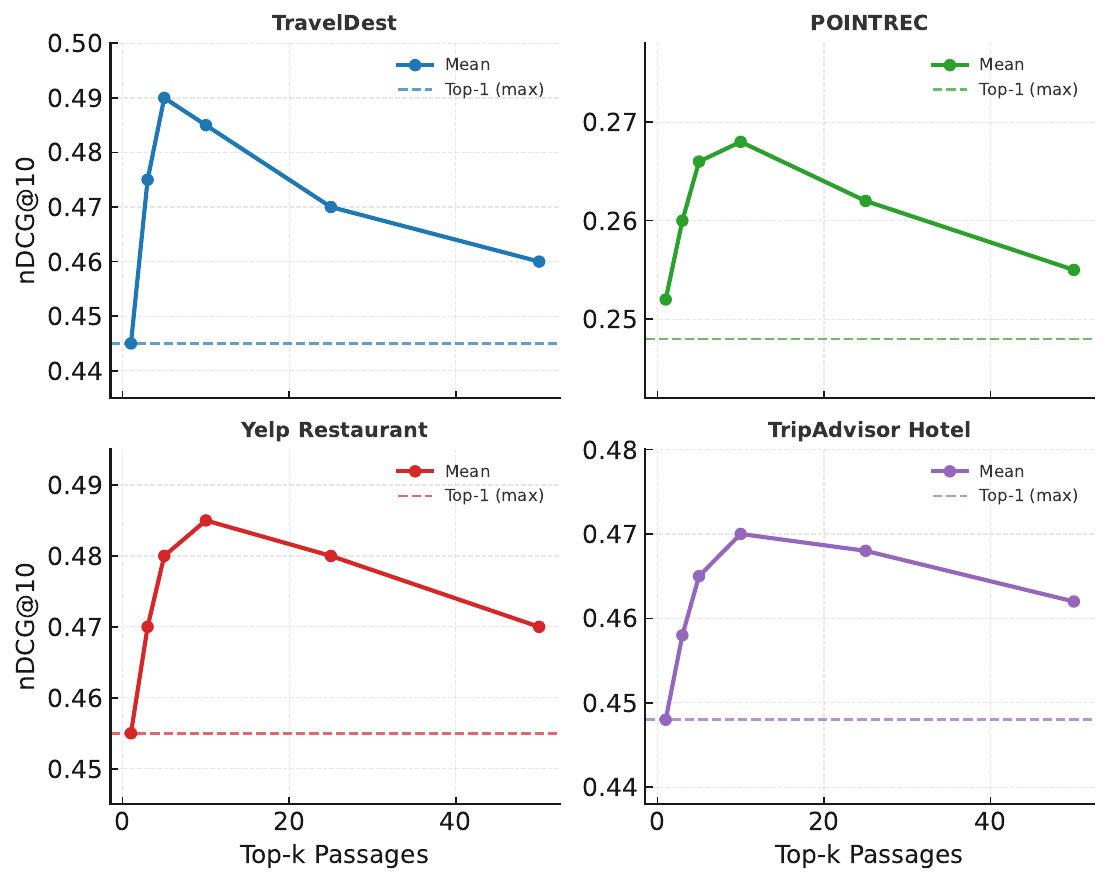}
    \caption{
    Performance of GPR-LLM under varying numbers of Top-$T$ passages and different aggregation functions for item-level relevance scoring.
    }
        \label{fig:travel_dest_topk}
\end{figure}

As shown in \autoref{fig:travel_dest_topk}, we can see that for mean aggregation method ndcg@10 increases when the number of top-k passages increases and achieves the best results when top-3 passages or top-10 passages are used. Then we observe a sharp decrease as more top-k passages are being aggregated. Our results suggest that a moderate choice of number of top-k passages (around 3-10) best balances these competing effects.

We compare different choices of aggregation functions to assess how the fusion of passage scores affects the final item recommendation quality. We test over [mean, max]. We can see that using mean as the aggregation function greatly outperforms the max method as we use more than one top-k passage for aggregation.

\end{document}